\documentclass[preprint,aps,tightenlines,showpacs]{revtex4}

\usepackage{graphicx}
\usepackage{bm}

\begin{document}


\title{\bf
An algebraic solution of the multichannel problem applied to low energy 
nucleon-nucleus scattering}
\author{K. Amos}
\email{amos@physics.unimelb.edu.au}
\affiliation{School of Physics, University of Melbourne,
Victoria 3010, Australia}

\author{L. Canton}
\email{luciano.canton@pd.infn.it}
\author{G. Pisent}
\email{gualtiero.pisent@pd.infn.it}
\affiliation{Istituto Nazionale di Fisica Nucleare, sezione di Padova, \\
e Dipartimento di Fisica dell'Universit$\grave {\rm a}$
di Padova, via Marzolo 8, Padova I-35131,
Italia}

\author{J. P. Svenne}
\email{svenne@physics.umanitoba.ca}
\affiliation{
Department of Physics and Astronomy,
University of Manitoba,
and Winnipeg Institute for Theoretical Physics,
Winnipeg, Manitoba, Canada R3T 2N2}

\author{D. van der Knijff}
\email{dirk@unimelb.edu.au}
\affiliation{Advanced Research Computing, Information Division,
University of Melbourne, Victoria 3010, Australia}

\date{today}

\begin{abstract}
Compound resonances in nucleon-nucleus scattering are related to the discrete 
spectrum of the target.  Such resonances can be studied in a unified and general
framework by a scattering model that uses sturmian expansions of postulated 
multichannel interactions between the colliding nuclei.  Associated with such 
expanded multichannel interactions are algebraic multichannel scattering 
matrices.  The matrix structure of the inherent Green functions not only 
facilitates extraction of the sub-threshold (compound nucleus) bound state 
spin-parity values and energies but also readily gives the energies and widths 
of resonances in the scattering regime.  
We exploited also the ability of the sturmian-expansion method to deal
with non-local interactions to take into account the strong non-local
effects introduced by the Pauli principle.  As an example, we have used the
collective model (to second order) to define a multichannel potential matrix 
for low energy neutron-$^{12}$C scattering allowing coupling between the 
$0^+_1$ (ground), $2^+_1$ (4.4389 MeV), and $0^+_2$ (7.64 MeV) states. The algebraic 
$S$ matrix for this system has been evaluated and the sub-threshold bound states
as well as cross sections and polarizations as functions of energy are 
predicted. The results are reflected in the actual measured data, and are shown 
to be consistent with expectations as may be based upon a shell model 
description of the target and of the compound nucleus.
\end{abstract}


\maketitle

\vskip -2.5cm

\section{Introduction}

Predicting the scattering of low energy (E $<$ 10 MeV) protons and neutrons from
nuclei is of wide-spread interest, as is their capture by same.  Intrinsically,
the probabilities of reactions (measured cross-section data) reflect the 
inherent structure and modes of excitation of the struck nucleus.  However, 
despite the large amount of data so far accrued, there remains much that has not
been measured, or cannot be, but the knowledge of which is prime input to many 
fields of study; fields as diverse as nuclear-astrophysics/cosmology and nuclear
stockpile stewardship.  

The most commonly studied process (experimentally and theoretically) is that of
elastic scattering; the outcome of such data analysis generally being required 
to begin studies of other reaction processes.  Typically, low-energy elastic 
scattering data show resonances upon a smoothly varying background, with 
cross-section magnitudes of the order of barns.  Resonances can be quite varied
in their character.  Particularly their widths (full widths at half maximum, 
FWHM) vary from a few eV to over an MeV.  For light mass targets the resonances
in the low energy regime tend to be distinct and but a few in number.  As mass 
increases, however, the numbers of resonances rapidly increase and the first 
tends lower in energy.  Many strongly overlap, so that a statistical approach 
to the analysis of scattering becomes feasible and utilitarian. But in the main,
data analysis to date has been phenomenological, with many parameter values
obtained by a fitting process. Little can be interpreted thereby about the 
intrinsic nature of the target, and the associated theory cannot be made 
predictive for estimations of the unmeasured (or unmeasurable) cross sections 
as needed in many other studies. That is also the case with many applications of
$R$-matrix theory of scattering~\cite{La66}, though at least one~\cite{Ri80}, 
for n-${}^{12}$C scattering, sought to specify the background scattering from 
an optical potential with partial widths of the resonances specified from 
($p$-$s$-$d$) basis shell-model wave functions of the nuclei involved, i.e. 
${}^{12,13}$C.

However, a more promising multichannel coupling theory of the scattering of 
neutrons from nuclei has been developed by the Padova 
group~\cite{Pi86,Ca87,Ca88,Ca90,Ca91,Pi95}; a theory based upon the work of 
Rawitscher and Delic~\cite{Ra82,Ra84}.  The approach is predicated upon making 
finite-rank separable representations of realistic interaction potentials and 
the properties of scattering matrices for such Schr\"{o}dinger interactions.
 
In brief, the approach starts with auxiliary sturmian functions (Weinberg 
states) forming a basis set in the interaction region. 
Initially one chooses a solvable potential problem
at any suitable fixed {\it negative} energy from which first generation 
sturmians can be 
specified in closed analytic form.  Second generation sturmians built upon the 
putative interaction potential matrices for a multichannel scattering problem of
interest then can be found as linear combinations of that first generation set.
The essential expansion coefficients result from a matrix diagonalization 
process.
 
The scheme enables expansions (usually truncated to finite rank for numerical
application) of the chosen interaction potentials in terms of those second 
generation sturmians, each in the form of a sum of separable interactions.  The 
analytic properties of the scattering matrix from a separable Schr\"{o}dinger 
potential gives the means by which a full algebraic solution of the multichannel
scattering problem can be realized.  Of note is that the algebraic structure of
the Green functions, that lie at the heart of the $S$ matrices, facilitates the
identification of all resonances.  The spin-parity ($J^\pi$), 
centroid, and width of each resonance can be ascertained without 
the need of a super-fine grid of energy values. It is important to 
identify especially narrow resonances that the scattering model predicts.

The multichannel formalism is outlined next in Sec.~\ref{multiT}. In this 
section there are two subsections, in the first of which we give some details of
separable representations of potential matrices in terms of sturmian functions. 
In the second subsection, the process by which we can identify and locate 
resonances is described.  The potential matrices specified 
using  the Tamura collective model of scattering~\cite{Ta65} then are discussed
in Sec.~\ref{tamura}. This section also outlines  how we deal with the Pauli
principle, using orthogonalizing pseudo-potentials.
 Results of calculations in which deformation of that 
collective model for interaction potential matrices has been taken to second 
order in the case of neutron scattering from ${}^{12}$C allowing coupling to the 
$0^+_1$ (ground), $2^+_1$ (4.4389 MeV), and $0^+_2$ (7.64 MeV) states, and found 
for neutron (laboratory) energies to 5 MeV, are presented and discussed in
Sec.~\ref{results}.  Note that we use laboratory frame energies when dealing 
with the scattering domain as the data to be used is so specified.  However,
when considering the bound states (below the n-$^{12}$C threshold) and/or 
spectra of $^{13}$C we use energies in the center of mass system as such
are used in tabulations of that spectra~\cite{Aj91}.

But the collective model of nuclear structure is limited in scope and eventually
we shall use this algebraic approach with microscopic (shell) model 
specifications of the target nuclei providing structure information which, when
folded with an effective two-nucleon force, will determine the base input 
nucleon-nucleus potential matrices.  That such shell model structures are 
reasonable for this purpose, at least with light mass targets, is considered 
finally in Sect.~\ref{shellmodel}.


\section{The multichannel $T$ matrix from separable interactions}
\label{multiT}

Consider a system of $C$ channels for each allowed scattering spin-parity 
$J^\pi$ with the index $c\ (=1,C)$ denoting the quantum numbers that identify 
each channel uniquely.  Suppose $c = 1$ designates the elastic channel.  The 
integral equation approach in momentum space for potential matrices 
$V_{cc'}^{J^\pi}(p,q)$, requires solution of coupled Lippmann-Schwinger (LS) 
equations giving a multichannel $T$ matrix of the form
\begin{eqnarray}
T_{cc'}^{J^\pi}(p,q;E) = V_{cc'}^{J^\pi}(p,q) &+& \mu \left[ 
\sum_{c'' = 1}^{\rm open}
\int_0^\infty V_{cc''}^{J^\pi}(p,x) \frac{x^2}{k^2_{c''} - x^2 + i\epsilon}
T_{c''c'}^{J^\pi}(x,q;E)\ dx \right.
\nonumber\\
&&\hspace*{0.5cm} \left.- \sum_{c'' = 1}^{\rm closed} \int_0^\infty
V_{cc''}^{J^\pi}(p,x) \frac{x^2}{h^2_{c''} + x^2} 
T_{c''c'}^{J^\pi}(x,q;E) \ dx \right]
\label{multiTeq}
\end{eqnarray}
Therein the open and closed channels contributions have been separated with the
respective channel wave numbers being 
\begin{equation}
k_c = \sqrt{\mu(E - \epsilon_c)}\hspace*{1.0cm} h_c = 
\sqrt{\mu(\epsilon_c - E)}\ ,
\label{wavenos}
\end{equation}
for $E > \epsilon_c$ and $E < \epsilon_c$ respectively with $\epsilon_c$ being 
the threshold energy of channel $c$. Here $\mu$ designates 
$2m_{\rm red}/\hbar^2$ with $m_{\rm red}$ being the reduced mass.  With the  
${J^\pi}$ superscript understood from now on, solutions of Eq.~(\ref{multiTeq})
are sought using expansions of the potential matrix elements in (finite) sums of
energy-independent separable terms,
\begin{equation}
V_{cc'}(p,q) \sim  \sum^N_{n = 1} {\hat \chi}_{cn}(p)\
\eta^{-1}_n\ {\hat \chi}_{c'n}(q)\ .
\label{finiteS}
\end{equation}
The link between the multichannel $T$ matrix and the scattering matrix
is~\cite{Ca91,Pi95}
\begin{eqnarray}
S_{cc'} &=& \delta_{cc'} -i \pi \mu \sqrt{k_c k_{c'}}\ T_{cc'}
\nonumber\\
&=& \delta_{cc'} - i^{l_{c'} - l_c +1} \pi \mu \sum_{n,n' = 1}^N \sqrt{k_c}
{\hat \chi}_{cn}(k_c) \left([\mbox{\boldmath $\eta$} - {\bf G}_0]^{-1}
\right)_{nn'}\ {\hat \chi}_{c'n'}(k_{c'})\sqrt{k_{c'}}\ ,
\label{multiS}
\end{eqnarray}
where now $c,c'$ refer to open channels only.  In this representation,
\textbf{${\bf G}_0$} and \mbox{\boldmath $\eta$} have matrix elements (for each
value of $J^\pi$ being understood)
\begin{eqnarray}
\left[{\bf{G}}_0 \right]_{nn'} &=& \mu\left[ \sum_{c = 1}^{\rm open} \int_0^\infty
{\hat \chi}_{cn}(x) \frac{x^2}{k_c^2 - x^2 + i\epsilon} {\hat 
\chi}_{cn'}(x)\ dx -
\sum_{c = 1}^{\rm closed} \int_0^\infty {\hat \chi}_{cn}(x) 
\frac{x^2}{h_c^2 + x^2}
{\hat \chi}_{cn'}(x)\ dx \right]
\nonumber\\
\left[{\mbox {\boldmath $\eta$ }}\right]_{nn'} &=& \eta_n\ \delta_{nn'}
\label{xiGels}
\end{eqnarray}
The bound states of the compound system are defined by the zeros of the matrix
determinant when the energy is $E < 0$ and so link to the zeros of
$\{ \left| \mbox{\boldmath $\eta$}-{\bf G}_0\right| \}$ when all channels in 
Eq.~(\ref{xiGels}) are closed.


\subsection{Sturmian expansion of a multichannel interaction}
\label{sturmexp}

It is convenient to define first generation sturmians as solutions of uncoupled
equations involving a known (local hermitian) potential matrix $U^{(0)}_{c}$,
\begin{equation}
G^{(0)}_{c}(E_{c}) U^{(0)}_{c} \left| \Phi^{(0)}_{{c} i}\right\rangle
= -\eta^{(0)}_{{c} i} \left| \Phi^{(0)}_{{c} i}\right\rangle\ ,
\label{first_gen}
\end{equation}
where $G^{(0)}_c(E_c)$ is the free Green function evaluated at any
suitable (arbitrary) {\it negative} energy $E_c$.  In these studies
we chose 
\begin{equation}
E_c \equiv B = -1 MeV\ ,
\label{new_E}
\end{equation}
which is independent of the channel.
Then, as both operators $G^{(0)}_c$ and $U^{(0)}_c$ are hermitian,
\begin{equation}
\left\langle \Phi^{(0)}_{{c} i} \right| U^{(0)}_{c} G^{(0)}_{c}(E_{c})
= -\eta^{(0)}_{{c} i} \left\langle \Phi^{(0)}_{{c} i}\right|\  .
\label{first_cc}
\end{equation}
These eigenfunctions satisfy a potential orthonormality relation,
\begin{equation}
\langle \Phi^{(0)}_{{c} i} \left| U^{(0)}_{c} \right| \Phi^{(0)}_{{c} j}\rangle
= \eta^{(0)}_{{c} i} \delta_{ij}\ ,
\label{orthonorm}
\end{equation}
where the normalization has been chosen to be consistent with a potential 
completeness relation of
\begin{equation}
\sum_{i=1}^\infty U^{(0)}_{c} \left| \Phi^{(0)}_{{c} i}\right\rangle
\left[\eta^{(0)}_{{c} i} \right]^{-1} \left\langle \Phi^{(0)}_{{c} i} \right|
U^{(0)}_{c}
= U^{(0)}_{c} \ .
\label{complete}
\end{equation}
A negative sign appears in Eqs.~(\ref{first_gen}) and (\ref{first_cc}) so 
that the orthonormality and completeness relations are consistent with the 
convention that $\left|\Phi^{(0)}_{{c} i}\rangle\right.$ are purely real 
functions.  Details of how first generation sturmians may be evaluated have been
published~\cite{Ca88}, but are given in brief in Appendix~\ref{stur-gen}
for completeness.

Assuming that the infinite sums in the expansions
Eq.~(\ref{complete}) can be truncated at a number $N_1$ of terms, sufficiently
large that all important elements of the actual multichannel scattering 
potential matrices can be expressed equivalently by either of two separable 
approximations, one finds
\begin{eqnarray}
V_{{c} {c'}} &\sim& \sum_{i=1}^{N_1} U^{(0)}_{c} \left| \Phi^{(0)}_{{c} i}
\right\rangle \left[\eta^{(0)}_{{c} i} \right]^{-1} \left\langle 
\Phi^{(0)}_{{c} i}
\right| V_{{c} {c'}}
\sim \sum_{i=1}^{N_1} V_{{c} {c'}} \left| \Phi^{(0)}_{{c'} i}\right\rangle
\left[\eta^{(0)}_{{c'} i} \right]^{-1} \left\langle \Phi^{(0)}_{{c'} i} \right|
U^{(0)}_{{c'}}
\nonumber\\
&\equiv&\ V^{(1)}_{{c} {c'}}\ .
\label{Vab_sep}
\end{eqnarray}

The procedure is to introduce second generation sturmians as eigenvectors of the
coupled-channel homogeneous equations involving this first generation 
approximation $V^{(1)}_{{c} {c'}}$,
\begin{equation}
\sum^{\Gamma}_{{c'} = 1} G^{(0)}_{{c}} V^{(1)}_{{c} {c'}} \left|
\Phi^{(1)}_{{c'} p}\right\rangle = - \eta^{(1)}_p \left| \Phi^{(1)}_{{c} p}
\right\rangle\ ,
\label{second_gen}
\end{equation}
where now it is assumed that the channel numbers are finite in extent 
($\Gamma$).  With the right side form of the first generation approximation for 
$V^{(1)}_{cc'}$, the expansion of the second generation basis in terms of the 
first is
\begin{equation}
\left| \Phi^{(1)}_{{c} p}\right\rangle = \sum^{N_1}_{j = 1} Q_{{c} j, p}
\left| \Phi^{(0)}_{{c} j}\right\rangle\ ,
\label{second_first}
\end{equation}
where the coefficients are given by
\begin{equation}
Q_{{c} j, p} = \sum_{{c'}} \left[\eta^{(1)}_p\right]^{-1} \langle
\Phi^{(0)}_{{c} j} \left| V_{{c} {c'}} \right| \Phi^{(1)}_{{c'} p} \rangle\ .
\label{coeffs}
\end{equation}
These coefficients may be determined as solutions of a matrix equation that is 
formed by projecting the second generation sturmians given in 
Eq.~(\ref{second_first}) onto 
$\left\langle \Phi^{(0)}_{{c'} m}\right| V_{{c'} {c}}$ and summing 
over the channel index ${c}$ to find
\begin{equation}
\sum^{\Gamma}_{{c} = 1} \sum^{N_1}_{j = 1} \omega_{{c'} m, {c} j}\
Q_{{c} j,p} = \eta^{(1)}_p Q_{{c'} m, p}\ ,
\label{OmegaQ}
\end{equation}
where the $\omega$-matrix elements are
\begin{equation}
\omega_{{c'} m, {c} j} = \langle \Phi^{(0)}_{{c'} m} \left| V_{{c'} {c}}
\right| \Phi^{(0)}_{{c} j} \rangle\ .
\label{omega_mes}
\end{equation}

If the channel coupling problem is assumed to be fully described by the selected
set of $\Gamma$ channels involved, the potential matrix $V_{{c} {c'}}$ is 
hermitian. The diagonalizing matrix ${\bf Q}$ then can be chosen orthogonal, 
with the $\Gamma \times N_1$ eigenvalues $\eta^{(1)}_p$ being purely real 
quantities.  Then, a completeness relation~\cite{Ca87,Ca88,Ca90} for the 
potential $V_{cc'}$ can be established, namely
\begin{eqnarray}
V_{{c} {c'}} &\cong& \sum_{p=1}^{\Gamma \times N_1} \sum_{{c''} = 1}^{\Gamma}
\sum_{c''' = 1}^{\Gamma} V_{{c} {c''}} \left| \Phi^{(1)}_{{c''} p} 
\rangle \right.
\left[ \eta^{(1)}_p\right]^{-1} \left . \langle \Phi^{(1)}_{c''' p}\right|
V_{c''' {c'}}
\nonumber\\
&=& \sum^{\Gamma \times N_1}_{p = 1} \left| \chi_{{c} p} \rangle \right.
\frac{1}{\eta^{(1)}_p} \left. \langle \chi_{{c'} p}\right|
\equiv V^{(2)}_{{c} {c'}}\ ,
\label{V_sep}
\end{eqnarray}
which provides the separable expansion of the potential matrix in terms of 
second generation (coupled-channel) sturmians. The form factors defined by
\begin{equation}
\left| {\chi_{{c} p}} \rangle \right.= \sum^{\Gamma}_{c' = 1} V_{{c} {c'}}
\left| \Phi^{(1)}_{{c'} p} \rangle \right.\ ,
\label{chi_facts}
\end{equation}
have been discussed in detail elsewhere~\cite{Ca87,Ca88,Ca90,Ca91,Pi95}.

With  $\Phi_{cn}(r)$, the Weinberg (or sturmian) functions in coordinate space,
those form factors then are given by
\begin{equation}
\chi_{cp}(r) = \sum^\Gamma_{c'=1} \int_0^\infty V_{cc'}(r r') 
\Phi^{(1)}_{c'p}(r')\ dr'\ ,
\label{radialFs}
\end{equation}
for a nonlocal potential  matrix $V_{cc'}(r'r)$, while for a local interaction 
they have the form
\begin{equation}
\chi_{cn}(r) = \sum^\Gamma_{c'=1} V_{cc'}(r) \Phi^{(1)}_{c'n}(r)\ .
\label{localFs}
\end{equation}
The form factors used to define the $T$ and $S$ matrices in
Eqs.~(\ref{multiTeq}) and (\ref{multiS}) however are in momentum space.  
Those momentum space form factors are the  Fourier-Bessel transforms
\begin{equation}
{\hat \chi}_{cn}(p) = \left[ \frac{2}{\pi} \right]^{\frac{1}{2}} \frac{1}{p}
\int_0^\infty F_{\ell}(pr) \chi_{cn}(r)\ dr\ ,
\label{FBtrans}
\end{equation}
where $\ell$ is the orbital angular momentum quantum number.

In application, it may be feasible to reduce the dimensionality of the problem 
by truncating the above to have $N_2 < (\Gamma \times N_1)$. The retained terms
still are combinations of all of the $\Gamma \times N_1$ first generation 
sturmians however. Thus all relevant features of the fuller calculation can be 
included in a smaller basis calculation. The choice may reside with the listing 
of the eigenvalues in an ordered (decreasing magnitude) set. Also, note that 
the extension to higher generation expansions has been 
investigated~\cite{Ca87,Ca88,Ca90}, but no significant improvement in the form 
developed with the second generation method was found.

Formally, a set of coupled equations can be replaced by one for just the elastic
channel alone from which one can define, in configuration space, the optical 
potential for elastic scattering.  Of note is that, even assuming a local form 
for the elastic channel element of the potential matrix, the resulting optical 
potential will be energy dependent and nonlocal~\cite{Ca87,Ca88,Ca90,Ca91a}.
For completeness, a brief development of the optical potential resulting from 
the sturmian expansion method is given in Appendix~\ref{Append-one}.

\subsection{Resonance identification}

In this section we describe a numerical technique for a rapid determination of 
all narrow resonances (in nucleon scattering from spin zero targets) arising 
from a system of coupled-channel Schr\"{o}dinger equations. We consider only the
elastic scattering channel for which the scattering matrix is recast (for each 
$J^\pi$ and with $k = k_1$) using trivial matrix manipulation of 
Eq.(~\ref{multiS}).
\begin{eqnarray}
S_{11} &=& 1 - i \pi \mu \sum_{nn'=1}^N k \ {\hat \chi}_{1n}(k)
\left[\left(\mbox{\boldmath $\eta$} -
{\bf G}_0\right)^{-1}\right]_{nn'}{\hat \chi}_{1n'}(k)
\nonumber\\
&=& 1 - i \pi \mu \sum_{nn'=1}^N k \ {\hat \chi}_{1n}(k) {1\over \sqrt{\eta_n}}
\left[\left({\bf 1} -
\mbox{\boldmath $\eta$}^{-{1\over 2}}{\bf G}_0\mbox{\boldmath 
$\eta$}^{-{1\over 2}}
\right)^{-1}\right]_{nn'} {1\over \sqrt{\eta_{n'}}} {\hat \chi}_{1n'}(k)\, .
\label{x11}
\end{eqnarray}
Here, the diagonal (complex) matrix $\mbox{\boldmath $\eta$}^{-{1\over 2}}$ is 
defined as
\begin{equation}
\left[\mbox{\boldmath $\eta$}^{-{1\over 2}}\right]_{nn'}
= \delta_{nn'}\ {1\over \sqrt{\eta_n}} \, .
\end{equation}
The complex-symmetric matrix
$\mbox{\boldmath $\eta$}^{\frac{1}{2}}{\bf G}_0\mbox{\boldmath 
$\eta$}^{-{1\over 2}}$ is then diagonalized
\begin{equation}
\sum_{n'=1}^N
{\eta_n}^{-{1\over 2}}\left[{\bf G}_0\right]_{nn'}
{\eta_{n'}}^{-{1\over 2}} \tilde{Q}_{n'r} = \zeta_r \tilde{Q}_{nr} \, ,
\label{eigen}
\end{equation}
and the evolution of the complex eigenvalues $\zeta_r$ with respect to energy 
define resonance attributes.  Resonant behavior occurs when one of the complex 
$\zeta_r$ eigenvalues passes close to the point (1,0) in the Gauss plane.  From 
Eq.~(\ref{x11}) it is evident that the S-matrix has a pole structure at the 
corresponding energy where one of these eigenvalues approach unity, since one 
can write
\begin{equation}
\left[\left({\bf 1} -  \mbox{\boldmath $\eta$}^{-{1\over 2}}
{\bf G}_0\mbox{\boldmath $\eta$}^{-{1\over 2}} \right)^{-1}\right]_{nn'}
=\sum_{r=1}^N \tilde{Q}_{nr}{1\over 1-\zeta_r}\tilde{Q}_{n'r} \, .
\end{equation}
These we designate as the resonance identifier equations in the text to follow.

The eigenvalues $\zeta_r$ correspond to the positive-energy eigenvalues of the
homogeneous (sturmian) problem with potential $V^{(2)}_{cc'}(p,q)$ as given by
Eqs.~(\ref{finiteS}) and (\ref{V_sep}).  In operator form the relevant 
sturmian equations are now
\begin{equation}
\sum_{c'}G_c^{(0)}(E) V^{(2)}_{cc'}|\phi^{(1)}_{c'r}(E)\rangle
=\bar\zeta_r(E) |\phi^{(1)}_{cr}(E)\rangle \, .
\label{Stur-res}
\end{equation}
Note that here the sign convention is different with respect to
Eqs.~(\ref{first_gen}) and (\ref{first_cc}) so that the attractive eigenvalues
will range in the upper-half Gauss plane as is conventionally used for 
resonance identifications.

To show that $\bar\zeta_r(E)$ and $\zeta_r(E)$ are the same, multiply both sides
of Eq.~(\ref{Stur-res}) by the potential $V^{(2)}_{cc'}$, sum over the relevant 
channels, and  use  Eq.~(\ref{V_sep}) for $V^{(2)}_{cc'}$.  The result is
\begin{equation}
\sum_{n,n'=1}^N |\chi_{cn}\rangle {1\over \eta_n} \left[ G_{0}(E)\right]_{nn'}
{1\over \eta_{n'}} \left(\sum_{c'}\langle 
\chi_{c'n'}|\phi^{(1)}_{c'r}(E)\rangle\right)
= \bar\zeta_r(E) \sum_{n=1}^N |\chi_{cn}\rangle {1\over \eta_n}
\left(\sum_{c}\langle \chi_{cn}|\phi^{(1)}_{cr}(E)\rangle\right) \, .
\end{equation}
Projection onto the bi-orthogonal states then yields
\begin{equation}
\sum_{n'=1}^N {1\over {\eta_n}} \left[ G_{0}(E)\right]_{nn'} {1\over 
{\eta_{n'}}}
\left(\sum_{c'}\langle \chi_{c'n'}|\phi^{(1)}_{c'r}(E)\rangle\right)
= \bar\zeta_r(E) {1\over \eta_n}
\left(\sum_{c'}\langle \chi_{c'n}|\phi^{(1)}_{c'r}(E)\rangle\right) \, ,
\end{equation}
which is equivalent to Eq.~(\ref{eigen}) provided that one make the 
identifications
\begin{equation}
\zeta_r(E) = \bar\zeta_r(E) \hspace*{0.5cm} ; \hspace*{0.5cm}
\tilde Q_{nr} = {1\over \sqrt{\eta_{n}}}
\left(\sum_{c'}\langle \chi_{c'n}|\phi^{(1)}_{c'r}(E)\rangle\right)\, .
\end{equation}
Thus the eigenvalues $\zeta_r$ of Eq.~(\ref{eigen}) are the positive-energy 
sturmian eigenvalues of the potential $V^{(2)}_{cc'}(p,q)$.  Their general 
properties are well known (see, $e.g.$, Ref.~\cite{We65}) and 
they can be evaluated reliably~\cite{Ra91}.


\section{A collective model of the potential matrices}
\label{tamura}

The basic scattering model used was that for collective excitations as defined 
by Tamura~\cite{Ta65}.  Such has been  
used before with some success to study resonance scattering~\cite{Pi67,Pa69}, 
and thus is our choice for this study. However, we allow
certain extensions to the usual collective model specification.
Specifically, all terms
to order `$\beta^2$' in deformation have been carried given that the collectivity of
the nucleus studied is strong. Also, the n-$^{12}$C potential field was allowed 
to  have central ($V_0$), $\ell^2$-dependent ($V_{\ell \ell}$), and spin-orbit 
($V_{\ell s}$) components.  And, as suggested in Ref.~\cite{Mi71}, a spin-spin 
term ($V_{s s}$) has also been included in the central components.

The basis of channel states is defined by the coupling
\begin{equation}
\left| c \right\rangle = \left| ({\ell}s) {\cal J} I J^\pi \right\rangle =
\biggl[
\left| ({\ell} \otimes s)_{\cal J}\right) \otimes \left| \psi_I^{(\alpha)}
\right\rangle
\biggr]_J^{M,\pi}\ ,
\end{equation}
where consideration is restricted to $s= \frac{1}{2}$, and the target states are
denoted by $\left| \psi_I^{(\alpha)} \right\rangle$ ($\alpha$ denotes any other
quantum numbers necessary to uniquely define the target states).  In coordinate
space, the $\left| (l \otimes s)_{\cal J}\right)$ are the spin-angle functions 
of the relative motion wave function.

\subsection{The channel-coupling potential matrix}

Again with each $J^\pi$ hereafter understood, and by disregarding deformation 
temporarily, the (nucleon-nucleus) potential matrices may be written
\begin{eqnarray}
V_{c'c}(r) &=& \langle (\ell's) {\cal J'} I' \left|\ W(r)\ \right|
(\ell s) {\cal J} I \rangle
\nonumber\\
&=& f(r) \biggl\{ V_0 \delta_{c'c} + V_{ll} [ {\bf {\ell \cdot 
\ell}} ]_{c'c}
+ V_{ss} [ {\bf {s \cdot I}} ]_{c'c} \biggr\} + g(r) V_{ls}
[{\bf {\ell \cdot s}}]_{c'c}\ ,
\label{www}
\end{eqnarray}
in which local form factors have been assumed.  This potential matrix form 
results when one considers the basic Tamura collective model~\cite{Ta65} 
typically with  Woods-Saxon form factors,
\begin{equation}
f(r) = \left[1 + e^{\left( {{r-R}\over a} \right)} \right]^{-1}
\hspace*{0.3cm} ; \hspace*{0.3cm} g(r) = \frac{1}{r} \frac{df(r)}{dr}.
\end{equation}
Deformation then is included by means of the rotational model~\cite{Ta65} 
approach in which the nuclear surface is defined by
\begin{equation}
R = R_0 ( 1 + \epsilon) \hspace*{0.5cm} ; \hspace*{0.5cm}
\epsilon = \sum_{L (\ge 2)} \sqrt{\frac {4 \pi}{2L+1}}\  \beta_L \
[{\bf Y}_L (\hat r) {\bf \cdot} {\bf Y}_L(\hat \Upsilon)]\ ,
\label{CollRad}
\end{equation}
where $\hat\Upsilon$ designates internal target coordinates.  Expanding $f(r)$ 
to order $\epsilon^2$ gives
\begin{equation}
f(r) = f_0(r) + \epsilon \left[\frac {df(r)}{d \epsilon}\right]_0
+ \frac {1}{2} \epsilon^2 \left[ \frac {d^2 f(r)}{d \epsilon^2}\right]_0\ .
\end{equation}
There is a similar equation for $g(r)$. Both $f(r)$ and $g(r)$ are now radial 
operators.

In what follows that extended (collective) model is developed assuming that only
one multipole $L$ need be retained in Eq.~(\ref{CollRad}); the definition of 
$\epsilon$. For the example we give later, of low energy neutron scattering from
${}^{12}$C, that value is 2. However it is useful to specify the potential 
matrices for general $L$, as the results form a basis from which to develop the 
theory when more than one multipole is needed. 

Using the property of tensor products ~\cite{Va88},
\begin{equation}
[{\bf Y}_L(\hat r) {\bf \cdot} {\bf Y}_L(\hat \Upsilon)]^2 = \frac {1}{4 \pi}
\sum_{\ell\ {\rm even}}^{2L} \frac {(2L+1)^2}{2\ell +1}\ 
\langle L 0 L 0 \vert \ell 0
\rangle^2\ [{\bf Y}_{\ell}(\hat r) {\bf \cdot} {\bf Y}_{\ell}(\hat 
\Upsilon)]\ ,
\end{equation}
the radial operators are given by
\begin{eqnarray}
f(r) =  f_0(r) &+& \sqrt{\frac{4\pi}{2L+1}} \beta_L
\left[ {\bf Y}_L {\bf \cdot} {\bf Y}_L \right] \frac{df_0(r)}{dr}
\nonumber\\
&+& \frac{1}{2} \beta^2_L (2L+1) \sum^{2L}_{\ell\ {\rm even}} 
\frac{1}{(2\ell+1)}
\left\langle L 0 L 0 \vert \ell 0 \right\rangle^2
\left[ {\bf Y}_\ell  {\bf \cdot} {\bf Y}_\ell \right] \frac{d^2f_0(r)}{dr^2}\ .
\label{fff}
\end{eqnarray}
A similar equation applies to $g(r)$.

To determine the deformed channel potential, it is not simply a matter of taking
the matrix elements of the radial operators between channels states $c$ and $c'$
and substituting them into Eq.~(\ref{www}).  The channel potential expression 
involves matrix elements of the products of two operators and so one must first
make symmetric the potential matrix form and use a suitable completeness
relations of the type $\sum_c|c><c|=1$ to separate the action of product 
operators.  The correct form of the matrix $V_{c'c}(r)$ is
\begin{eqnarray}
V_{c'c}(r) = V_0 f_{c'c}(r) &+& \frac {1}{2} V_{ll}  \sum_{c''} \left\{
[{\bf \ell} {\bf \cdot} {\bf \ell}]_{c'c''} f_{c''c}(r)
+ f_{c'c''}(r) [{\bf \ell} {\bf  \cdot} {\bf \ell}]_{c''c} \right\}
\nonumber \\
&+& \frac {1}{2} W_{ls}  \sum_{c''} \left\{
[{\bf \ell} {\bf \cdot} {\bf s}]_{c'c''}g_{c''c}(r)
+ g_{c'c''}(r) [{\bf \ell} {\bf \cdot} {\bf s}]_{c''c} \right\}
\nonumber \\
&+& \frac {1}{2} V_{ss}  \sum_{c''} \left\{ [{\bf s} {\bf \cdot} {\bf 
I}]_{c'c''}
f_{c''c}(r) + f_{c'c''}(r) [{\bf s} {\bf  \cdot} {\bf I}]_{c''c} \right\}\ ,
\label{potchan2}
\end{eqnarray}
where $ W_{ls} = 2 V_{ls} \lambda_\pi^2$.  Using the Woods-Saxon forms, details
of these matrix elements are given in Appendix~\ref{Append-two}.

A feature of our calculations is that the deformation is taken in all terms of 
the potential.  In Ref.~\cite{Ta65}, only the central part of the potential was
taken to be deformed.  This is an important point. It is incorrect to disregard 
the effect of deformation on the non-central parts of the potential, 
particularly so if large deformation values are to be used.

\subsection{The Pauli exclusion principle}

The standard coupled-channel approach to nucleon-nucleus scattering involves 
elastic and inelastic processes between two particles where at least one of the
two is itself composite and interacts with the other through effective (optical)
potentials.  One must keep in mind, however, that the underlying process is a 
complicated many-body scattering problem which requires consideration of the 
Pauli exclusion principle.

It is a known fact that phenomenological (coupled-channel) models coupling
collective deformations with single-particle optical potentials violate the
Pauli exclusion principle~\cite{MW69}.  In single-channel scattering processes
this violation does not represent a severe problem because the exclusion 
principle can be taken into account in the scattering process naturally if the
single-particle potential contains a series of deep bound states (the forbidden
states) to which the scattering wave function is orthogonal by construction.

In a coupled-channel model another problem arises because, in addition to the 
orthogonality of the scattering wave function to a series of forbidden states in
the elastic channel, one has to eliminate also all virtual transitions to the 
forbidden states in the excited channels.  A failure to do so implies the 
construction of an erroneous scattering wave function; one which would be 
unavoidably contaminated by the unphysical couplings to states of the 
nucleon-(excited nucleus) system forbidden by the exclusion principle.

The main effect of the Pauli principle is well represented by the suppression of
a part of the phase space which would be otherwise accessible to the system. 
This implies that the two fragments being scattered must negotiate between 
themselves with effective potentials where certain states forbidden by the Pauli
principle are excluded explicitly.

An efficient method to achieve this goal is obtained by introducing the 
orthogonalizing pseudo-potentials (OPP)~\cite{Ku78,Kr74} The OPP method is a 
variant of the orthogonality condition model by Saito~\cite{Sa69} which 
allows projection of the scattering solution onto the subspace permitted by the
exclusion principle through a suitable renormalization of the two-particle
interaction matrices. All virtual transitions to the forbidden states are 
therefore eliminated by the redefinition of the two-body interaction operators.

Implementing this method we add an orthogonalizing pseudo-potential (OPP) term
to the collective potential matrices $V_{cc'}(r)$ giving
\begin{equation}  
{\cal V}_{cc'}(r,r') = V_{cc'}(r)\ \delta(r-r') + \lambda A_{c}(r) 
A_{c}(r')\ \delta_{c,c'}\ .
\end{equation}
The pseudo-potential is manifestly nonlocal in coordinate space and it is 
diagonal in
the channel index.  The function $A_c(r)$ is the radial part of the single 
particle wave function in channel $c$, spanning the phase-space excluded by the
Pauli principle in that particular channel. In the actual calculation, $A_c(r)$ 
is determined by solving the radial Schr\"odinger equation for each channel $c$ 
\begin{equation}
{d^2\over dr^2}A_c(r) + w_c(r)A_c(r)=0\ ,
\end{equation}
numerically with bound-state boundary conditions. Here the abbreviation
\begin{equation}
w_c(r)=\frac{2m}{\hbar^2}
\left[ E- V_{cc}(r) \right] - \frac{l(l+1)}{r^2} \, .
\end{equation}
has been used.  The OPP is added only for those channels $c$ containing single 
particle quantum numbers referring to closed shell configurations. For the
n-$^{12}$C system to be considered here, we eliminate all elastic and inelastic 
states related to the $0s_{\frac{1}{2}}$ and $0p_{\frac{3}{2}}$ configurations 
(deep bound states)
from the scattering equations.  Following the OPP method, the forbidden 
configurations are eliminated from the dynamical equation in the limit 
$\lambda \rightarrow \infty$.  For this study on low-energy nucleon-nucleus 
scattering, we have selected $\lambda \simeq 100 MeV$ as a value sufficiently 
large.

It is an important feature of the sturmian expansion method that one obtains an
algebraic solution for coupled-channel scattering for both local and nonlocal
interactions. The physical nonlocalities introduced by the Pauli principle are 
reflected in the strong nonlocal character of the OPP terms. Thus, we apply the 
sturmian expansion method to the Pauli corrected potential 
${\cal V}_{cc'}(r,r')$ and not to the original potential $V_{cc'}(r)$.
This leads to the following modifications of the sturmian expansion method:

1) The matrix $\omega_{c'm,cj}$ now has to be calculated according to
\begin{equation}
\omega_{{c'} m, {c} j} = \langle \Phi^{(0)}_{{c'} m} \left|
{\cal V}_{{c'} {c}}
\right| \Phi^{(0)}_{{c} j} \rangle\ ,
\label{omega_PAULI}
\end{equation}
however, the auxiliary sturmian base
$\left|\Phi^{(0)}_{{c'} m} \right\rangle $ can be maintained as defined before.

2) Diagonalizing the matrix $\omega$ leads to the coupled sturmian eigenstates
$\left| \Phi^{(1)}_{{c} p} \right\rangle$ and eigenvalues $\eta^{(1)}_p$,
thus providing the ingredients for the expansions of the potential matrices
\begin{eqnarray}
{\cal V}_{{c} {c'}} &\cong& \sum_{p=1}^{\Gamma \times N_1} 
\sum_{{c''} = 1}^{\Gamma}
\sum_{c''' = 1}^{\Gamma} {\cal V}_{{c} {c''}}
\left| \Phi^{(1)}_{{c''} p} \rangle \right.
\left[ \eta^{(1)}_p\right]^{-1} \left . \langle \Phi^{(1)}_{c''' p}\right|
{\cal V}_{c''' {c'}} = {\cal V}^{(2)}_{{c} {c'}}\
\nonumber\\
&=& \sum^{\Gamma \times N_1}_{p = 1} \left| \chi_{{c} p} \rangle \right.
\frac{1}{\eta^{(1)}_p} \left. \langle \chi_{{c'} p}\right|\ ,
\label{V_sep_PAULI}
\end{eqnarray} 
where the last equation implies
\begin{equation}
\left| {\chi_{{c} p}} \rangle \right.= \sum^{\Gamma}_{c = 1} {\cal 
V}_{{c} {c'}}
\left| \Phi^{(1)}_{{c'} p} \rangle \right.\ .
\label{chi_facts_PAULI}
\end{equation}
Therefore, the $\omega$-matrix is now calculated with the integrations 
\begin{eqnarray}
\omega_{{c'} m, {c} j} &=& \int_0^\infty \Phi^{(0)}_{{c'} m}(r)
{V}_{{c'} {c}}(r)
\Phi^{(0)}_{{c } j} (r) dr  \nonumber\\
&+&\delta_{cc'} \lambda
\left[\int_0^\infty A_c(r) \Phi^{(0)}_{{c} m} (r)  dr\right]
\left[\int_0^\infty A_c(r) \Phi^{(0)}_{{c} j} (r)  dr\right] \, ,
\label{omega_PAULI_r}
\end{eqnarray}
and the new potential form factors in coordinate space are given by
\begin{equation}
\chi_{cn}(r) = \sum^\Gamma_{c'=1} V_{cc'}(r) \Phi^{(1)}_{c'n}(r) 
+\lambda A_c(r)
\left[\int_0^\infty A_c(r') \Phi^{(1)}_{{c} n} (r')  dr'\right] \, .
\label{localFs_PAULI}
\end{equation} 

\section{Application: $n-^{12}$C scattering}
\label{results}

\begin{figure}
\includegraphics{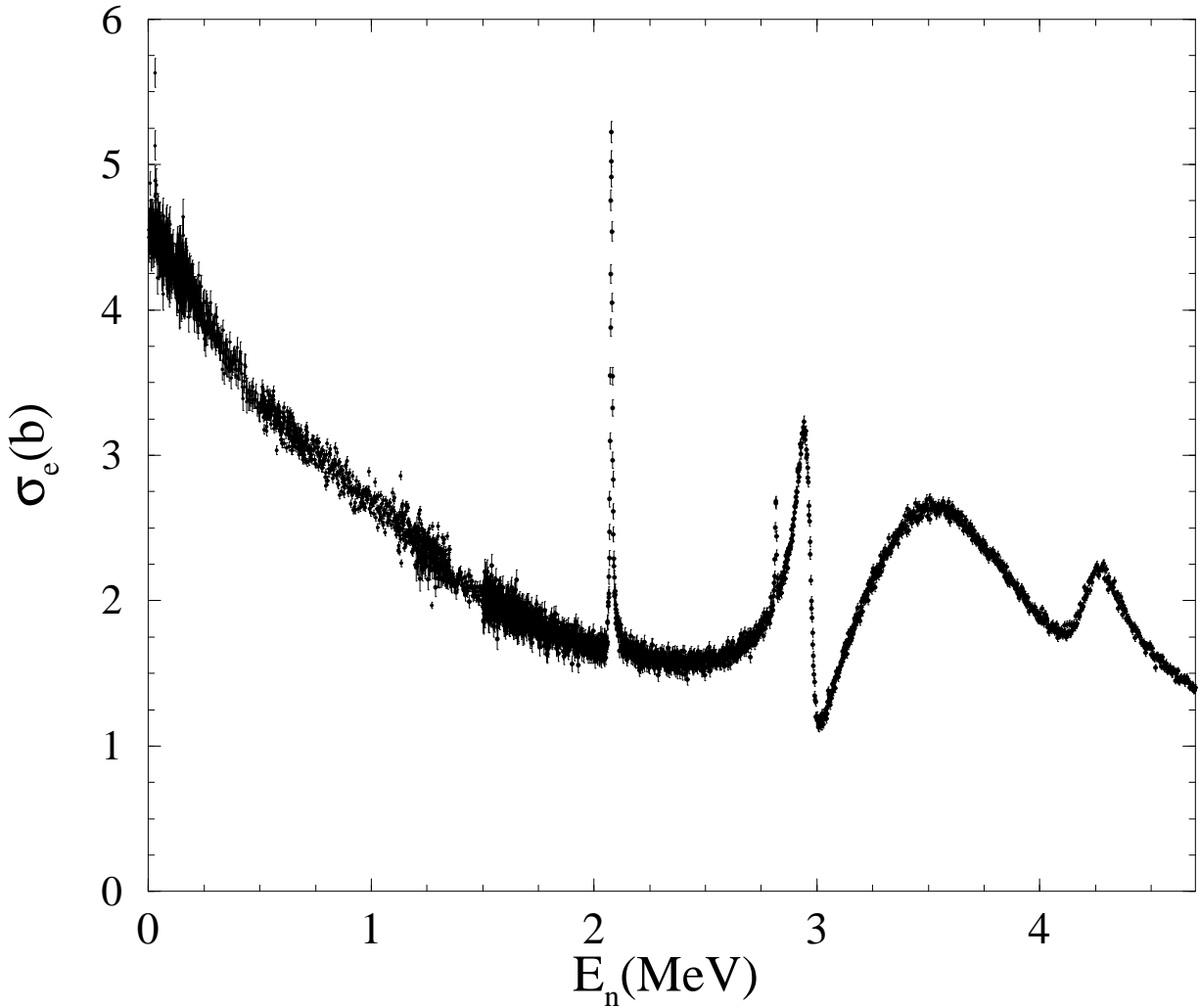}
\caption{ \label{C12data}
Measured elastic  cross sections for n-$^{12}$C scattering as functions
of neutron energy.}
\end{figure}

As a test case for study using the algebraic approach with the collective model
potential matrices, we consider neutron scattering from $^{12}$C. 
The measured cross section data are shown in Fig.~\ref{C12data}.  Those values were 
obtained using CINDA search in the web page of the National Nuclear Data Center,
Brookhaven ({\it www.nndc.bnl.gov}).  References for all the data sources are 
given therein.
The data set shown reveals very narrow and very broad resonances in the energy
range from 0 to 4 MeV. Most prominent of the narrow resonances are those at 
2.08 and at 2.75 MeV.  They have been assigned spin-parities of $\frac{5}{2}^+$
and $\frac{7}{2}^+$ respectively. The broad peaks have spin-parity assignments
of $\frac{3}{2}^+$, $\frac{3}{2}^+$, and  $\frac{1}{2}^-$ centered at 2.95, 
3.58, and 4.26 MeV respectively. Those resonances lie on a background that 
varies smoothly with energy from 4.6 barn.

In our analyses, three states 
in $^{12}$C have been taken as active. They are the $0^+_1$ (ground), the $2_1^+$ 
(4.4389 MeV), and the $0^+_2$ (7.96 MeV) states.  As the three states have the 
same ($+$) parity, the even and odd parity scattering channels then depend 
separately upon the even and odd parity input potentials respectively.  
To begin the search for suitable collective-model parameters,
we made the usual assumptions of the shell model,
and nuclear deformations; namely, that the nuclear radius is given by
$ R_0 = r_0A^{1/3}$, taking $ r_0 = 1.35\ {\rm fm}$, the diffuseness
as $ a_0 = 0.65\ {\rm fm}$, and that the depth of the potential be about
$ V_0 = -40.0\ {\rm MeV}$. The strength of the spin-orbit potential
was taken as $ V_{ls}^{(+)} =  6.7 \ {\rm MeV}$.
The value $\beta_2 = -0.6$
was taken to define the deformation of the ${}^{12}$C target nucleus.
As usual, the incoming nucleon is treated as a point particle.
The other parameters, the strengths of the $\ell^2$ and spin-spin
interactions, are not well known, and were used as fitting parameters.
As well, the other potential strengths and the deformation parameter were
considered as parameters that could be varied to improve the
representation of the data.
A first crude attempt to represent the experimental results
led us to increase the positive-parity central-potential strength to
$ V_0^{+} = -46.848\ {\rm MeV}$, set the $\ell^2$ term strength at
$ V_{ll}^{(+)} = 0.611 \ {\rm MeV}$ and both spin-spin term strengths at
$ V_{ss}^{(\pm)} =  -1.0 \ {\rm MeV}$.
The first generation
sturmians were obtained using a square well with parameters $\{B, R\} =
\{-1.0\ {\rm MeV }, 7.0\ {\rm fm} \}$
as the auxiliary potential.  Matrix sizes were limited to a 30 sturmian expansion
for each channel.  The resonance identifier equations as well as those for the
$S$ matrices were evaluated for all
channel spin-parities $J^\pi$ from ${\frac{1}{2}}^\pm$ to ${\frac{9}{2}}^\pm$.
Using this starting set of parameter values, by 
taking deformation through second order, lead to a very rich 
structure in the scattering and more importantly a structure that is reflected 
in measured data.

The scattering model we use does indeed
have the features seen in measured data though the centroids and widths can, 
and do, vary considerably with 
choice of parameter values.  Also, dependent upon the starting parameter set, 
there can be more narrow and/or weak resonances in this energy regime. However 
it is essential that a complete coupled-channel analysis be made of the 
scattering for the resonances to be considered sensible. Of course the 
deformation parameter $\beta_2$ has been assessed from diverse applications
of the collective model of excitation~\cite{Ho71} and in particular from the 
electron form factor and so should be rather constrained in any variation.
Likewise proton inelastic scattering cross sections from 
excitation of the 2$^+$ (4.4389 MeV) state indicate potential parameter
values that one may consider as ''sensible''. But those analyses, by and large, 
used the (distorted wave) Born approximation. Such an approximation might equally 
well be used in finding the background at low energies but is not 
appropriate when a study of the resonance attributes are to be made.

The end result of the search process we have made is the set of parameter values
\begin{equation}
\begin{array}{ll}
V_0^{(-)} = -49.1437\ {\rm MeV} & V_0^{(+)} = -47.5627 \ {\rm MeV}\\
V_{ll}^{(-)} = 4.5588\ {\rm MeV} & V_{ll}^{(+)} = 0.6098 \ {\rm MeV}\\
V_{ls}^{(-)} =  7.3836\ {\rm MeV} & V_{ls}^{(+)} =  9.1760 \ {\rm MeV}\\
V_{ss}^{(-)} =  -4.77\ {\rm MeV} & V_{ss}^{(+)} =  -0.052 \ {\rm MeV}\\
r_0 = 1.35\ {\rm fm} & \\
a_0 = 0.65\ {\rm fm} &  \beta_2 = -0.52\ . \\
\end{array}
\label{newParams}
\end{equation}

Initially the search process was extremely computer time consuming since, 
prior to the development (and use) of the resonance identifier equations, either 
the phases of the $S_{11}^{J^\pi}$ (elastic) scattering matrices and/or the 
calculated cross section for many energies to 5 MeV were used to specify the 
resonance energies and their FWHM.  But using the calculated cross section 
and/or the phase properties of the elastic channel $S$ matrices alone does not 
guarantee that weak and/or very narrow resonances will be evident. The choice of
energies and the step size used may not reveal characteristic effects that draw 
one's attention to the energy region for more intense study.  It is a hit or 
miss scenario.  Indeed with the initial coarse grid of energy steps of 0.01 MeV 
even the dominant ${{\frac{5}{2}}^+}$ resonance near  2.1 MeV MeV was not 
evident in the base calculations that were made.  If there were no other means 
by which resonance existence and centroid energies generally could be located,
predictions for poorly or as yet to be measured cross sections would have to be 
made using inordinately small energy steps over the whole range of interest; a 
major computing problem. Fortunately, the properties of the Green functions and 
of their eigenvalues enable the existence, number, and energy centroids of 
resonances to be found at the outset.  

Using the resonance identifier equations with sturmians built from the basic 
interaction potential, we obtained a sequence of bound 
states and resonance energies and widths for resonances in the neutron plus 
$^{12}$C system.  The actual values and how they are specified are presented in 
the second of the following subsections. In the first we illustrate the general 
behavior of the sturmian eigenvalues in Argand diagrams where the horizontal
axis gives the real part of the eigenvalue and the vertical, the corresponding 
imaginary part.

\subsubsection{Sturmian trajectories}

For each total angular momentum and parity, we have calculated the eigenvalues 
$\zeta_r(E)$ over the energy range $0.01$ to $4.96$ MeV with a constant step of 
$0.05$ MeV.  The eigenvalues in this energy regime, usually increase in magnitude 
with energy, with 
those coinciding with definable resonances in cross sections having values less
than $1+i0$ for energies below each associated resonance energy.  Thus even 
starting with a fairly coarse energy grid, as long as there are energy values 
both below and above the resonance centroid, no matter how narrow the resonance,
an eigenvalue will change from below to above $1 + i0$ for the two energy 
points in the grid that lie below and above that centroid.  Thus we learn from 
use of the resonance identifier equations not only how many resonances there are
in the energy range considered, but also where to make finer grid searches to 
better ascertain the characteristic properties of each resonance.
The results to be shown were all derived from the calculations made with
the potential matrices associated with the parameter set of Eq.~\ref{newParams}.
 
\begin{figure}
\includegraphics[angle=0,scale=0.7]{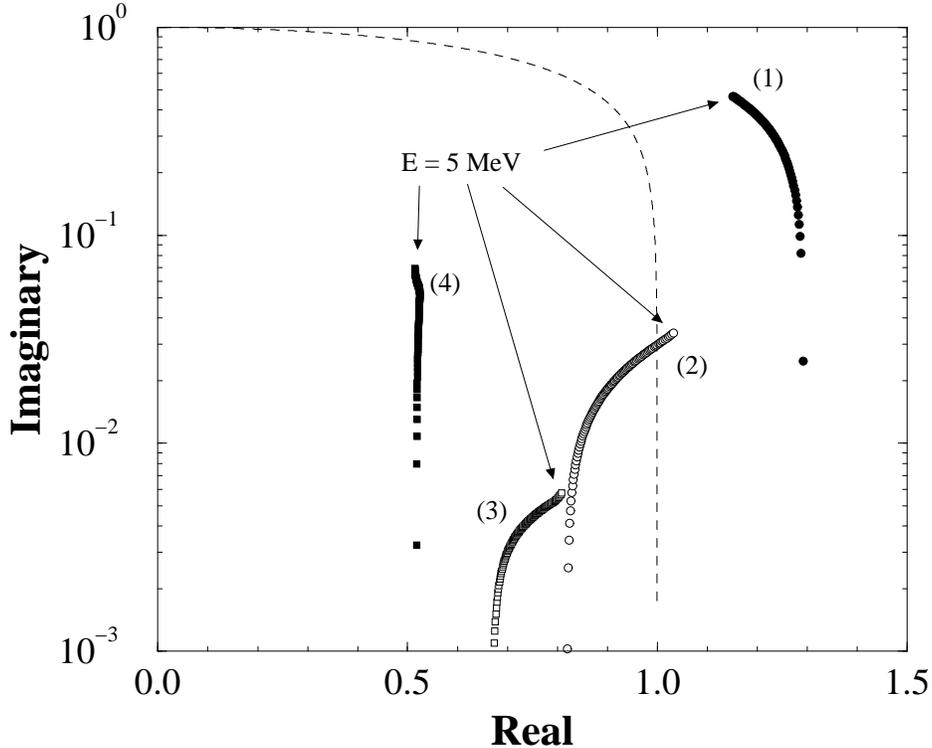}
\caption{\label{0.5plus} Argand diagrams for the energy 
variation of the sturmian eigenvalues for $J^\pi=1/2^+$.}
\end{figure}

In Fig.~\ref{0.5plus} 
we show the behavior of the four largest eigenvalues for $J^\pi={1\over 2}^+$ 
in the energy range to 5 MeV.  The unit circle is displayed by the dashed curve. 
Note that the graph is semi-logarithmic since all of these eigenvalues have  
small imaginary components. 
That is due to the elastic channel always being open.
Some trajectories continue below the graphed range. 
 All four at low energies have a variation that is
vertical to the real axis; a characteristic of the eigenfunction solutions for
$s$-waves.  The trajectories of two eigenvalues (curves (1) and (4)) 
show typical behavior of single-channel 
(potential) s-wave eigenvalues, as has been discussed in 
Ref.~\cite{We65}.  They do not correspond to any resonant structure.
The trajectories of the other two eigenvalues (curves (2) and (3)) shown in
Fig.~\ref{0.5plus} exhibit a quite different behavior; one which originates
from the coupled channel dynamics. 
In the energy range considered only that identified as (2) links to a 
resonance feature in scattering.

\begin{figure}
\includegraphics[angle=0,scale=0.7]{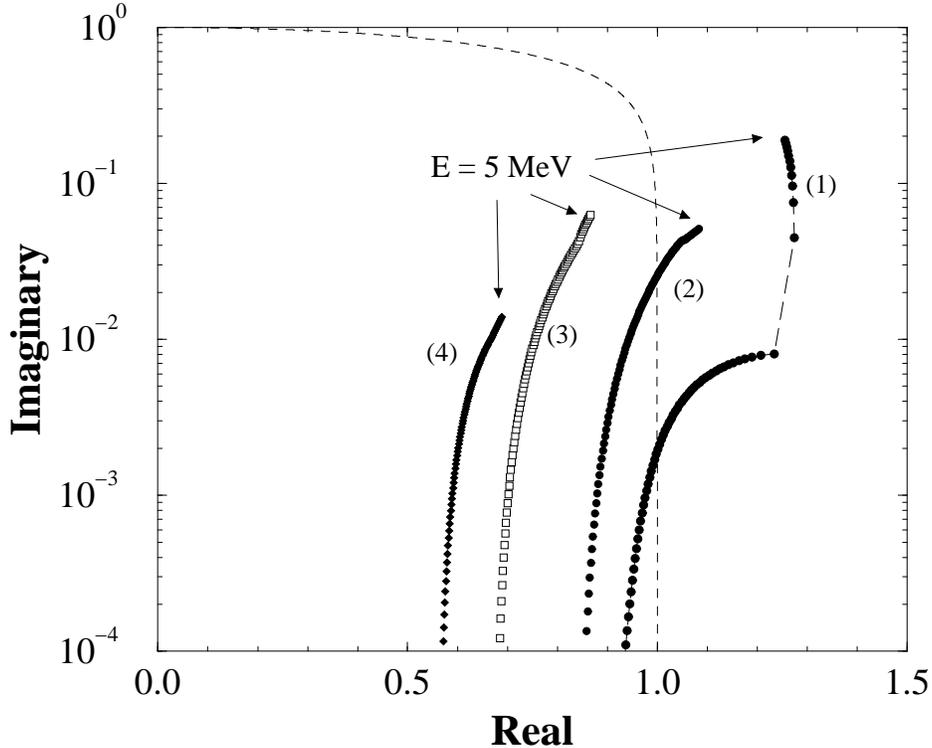}
\caption{\label{1.5plus} Argand diagram for the energy 
variation 
of the sturmian eigenvalues for $J^\pi=3/2^+$.}
\end{figure}

Next, the behavior of $J^\pi={3\over 2}^+$ eigenvalues are considered. The 
relevant Argand diagrams are given in Fig.~\ref{1.5plus}.
The point values of the largest eigenvalue (labelled (1)) has been i
connected by a
long-dashed line to guide the eye 
to link the energy sequence of the results. The actual trajectory is a cusp.
That is very evident with curve (1), but similar though more slight features
are evident in the trajectories (2) and (3).
That cusp feature of the eigenvalue trajectories links to the opening of
the $2^+_1$ state at 4.4389 MeV in the coupled-channel algebra.  

Note that these eigenvalues again have small imaginary parts and so the 
plot is semi-logarithmic and again some details of the trajectories lie
below the graphed range.  In this case the trajectories do not depart 
vertically from the real axis at the scattering threshold since they are 
$d$-wave solutions.  The (largest) eigenvalue 
clearly evolves well beyond the unit after crossing with a quite small 
imaginary part.  Thus it coincides with the lower energy $\frac{3}{2}^+$
compound  resonance in the cross section. Likewise the 
2$^{\rm nd}$ trajectory crosses the unit circle at higher energy 
and with a larger imaginary part. This coincides with the 
known second, broader, $\frac{3}{2}^+$ shape resonance at 3.4 MeV. 
The 3$^{\rm rd}$ and 4$^{\rm th}$ sturmian trajectories
shown in Fig.~\ref{1.5plus}  track towards the unit circle but have
not crossed before 5 MeV.

\begin{figure}
\includegraphics[angle=0,scale=0.7]{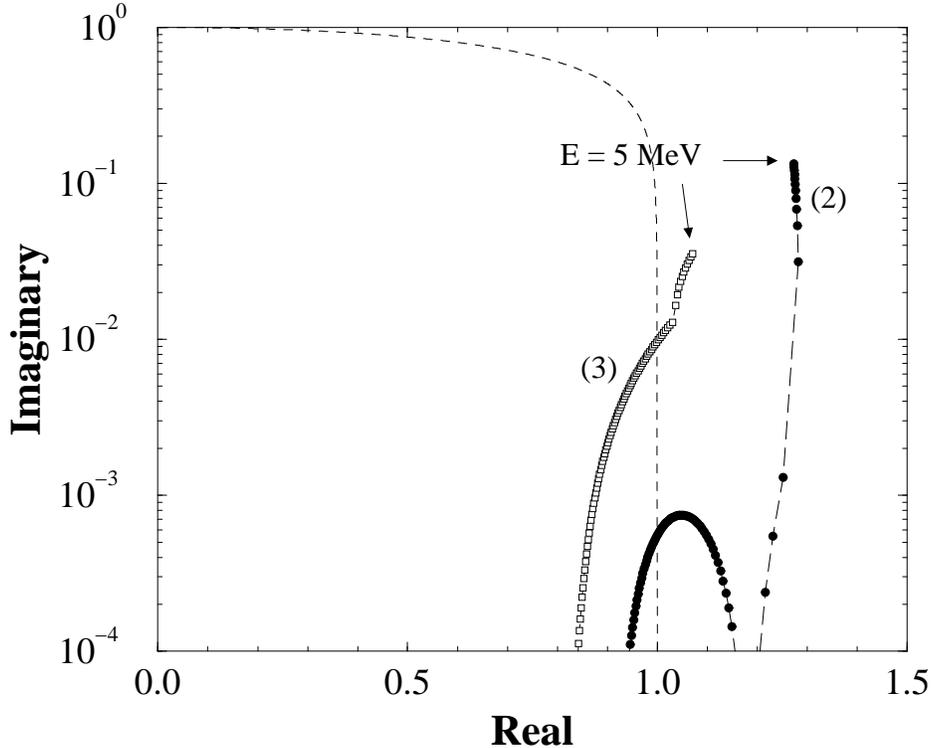}
\caption{\label{2.5plus-ir=2:3} Argand diagram for the energy 
variation of the sturmian eigenvalues for $J^\pi=5/2^+$.
}
\end{figure}

Finally, in Fig.~\ref{2.5plus-ir=2:3}, we show the Argand diagram of the 
2$^{\rm nd}$ and 3$^{\rm rd}$ largest eigenvalues for 
$J^\pi={5\over 2}^+$ since  
the first coincides with a state in the mass 13 spectrum below threshold.
The unit circle is represented therein by the dashed curve. 
Again the long-dashed line is simply to guide the eye to the energy variation 
of these results; results which give rise to a very narrow resonance at 
E$\simeq$2.086 MeV.  As the vertical scale again is logarithmic, 
the trajectories
continue below the graphed limit.  The trajectories also exhibit a non-vertical
''take-off' from the real axis commensurate with them portraying $d$-wave
eigenvalues.  Also both exhibit cusps which tag to the energy of the 
$2^+_1$ state.

\begin{figure}
\includegraphics[angle=0.0,scale=0.7]{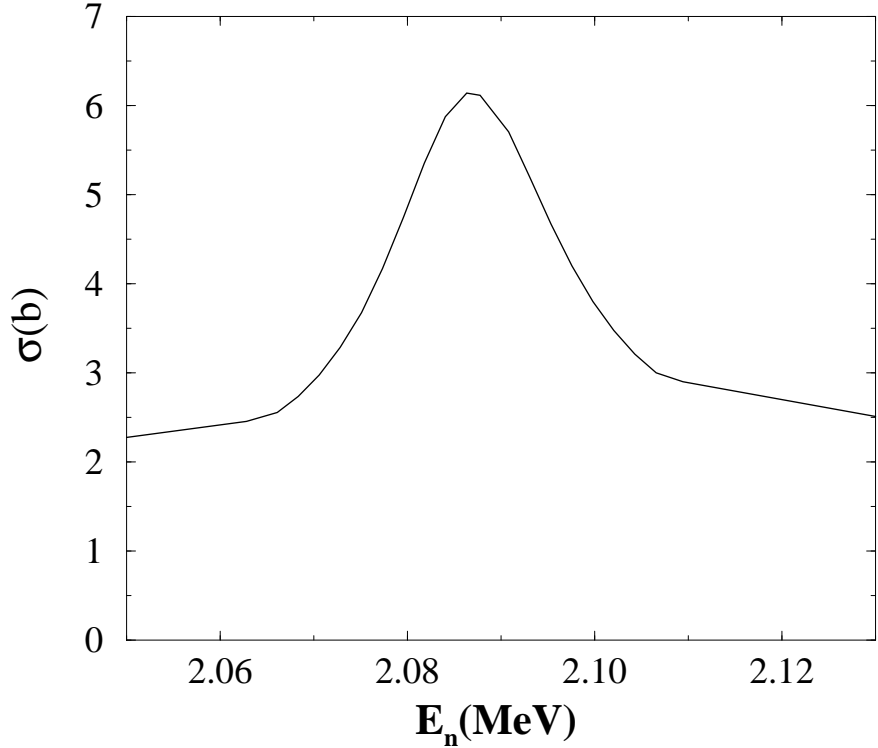}
\caption{\label{2.5plus-reson}
The $J^\pi = \frac{5}{2}^+$ resonance in the elastic scattering 
cross section.}
\end{figure}

Given that the imaginary component of the eigenvalue (curve (2)) is very
small, it is a very narrow resonance in the cross section, and the resonance 
centroid essentially is equivalent to  the energy at which
the real component of the eigenvalue itself is unity. 
That resonance in the total
cross section is shown in Fig.~\ref{2.5plus-reson}. It has a width of about 15
keV and a magnitude of over 6 barn.

\subsubsection{Calculation of resonance parameters}

To find the resonance centroids and widths, one first has to find all the 
energies where the real part of the sturmian eigenvalues crosses unity.  That 
determines an approximate value for the resonance energy $E_R$ by the condition
\begin{equation}
\zeta_r(E_R) = 1 + i\delta \, .
\label{e-res}
\end{equation}
Since these eigenvalues change smoothly with energy, it is possible to find 
these points with a rapidly converging predict-and-correct iterative procedure.
We have also estimated the width of these resonances, according to the 
approximation
\begin{equation}
{\Gamma_R} \simeq 2 \times {dE\over d[{\cal R}(\zeta_r)]} \delta
\label{g-res}
\end{equation}
where $d[{\cal R}(\zeta_r)]$ represents the differential of the real part of 
$\zeta_r$ at the resonance energy. 
These formulas (Eqs.~(\ref{e-res}) and (\ref{g-res})) are correct in the limit
$\delta << 1$, therefore they should be considered reliable only for the narrow
resonances. For the case of wider resonances, it is not difficult to find the 
more complicated expressions required. In the context of the present discussion,
however, they are not needed and indeed cross-section evaluations at reasonable
energy steps can be used to ascertain them quite easily.

\subsection{The theoretical elastic cross section and polarizations}
\label{Sec:Theocross}

The method described in the previous subsection enables us to determine with 
some precision, the centroid energies and widths of all the resonances. 
Similarly, we can find the energies of the bound states. At the same time we 
have sought a good theoretical description of the elastic cross section 
in compare with the data of Fig.~\ref{C12data}. The polarizations (at select 
scattering angles) will follow without having played any part in determining 
potential parameter values and so are ``predictions''.

Starting from the basic potential parameters, we 
determined all the resonance energies, widths and bound-state energies.
That set we denote as $\left\{ {E^{(0)}_i} \right\}$.  Then we changed each 
parameter by a small amount and calculated a matrix of 
(approximate) partial derivatives of the $E^{(0)}_i$ with respect to the 
parameters.  That matrix of derivatives we denote generically by 
$\left\{\frac{\partial E^{(0)}_i}{\partial \beta_p} \right\}$.  
Then, to first order, a new set of energies
$\left\{ E_i^{(1)}\right\}$ are given by the $n=1$ form of the recursive 
formula,
\begin{equation}
E_i^{(n)} \simeq E^{(n-1)}_i + \sum_p \frac{\partial 
E_i^{(n-1)}}{\partial \beta_p }
\Delta\beta_p
\label{linsys}
\end{equation}
where $\Delta\beta_p$ are the changes in parameters $\beta_p$ needed to produce
the new $E_i^{(n)}$.  If the number of parameters is equal to the number of 
energy values, this set of equations can be considered as a linear system to 
solve for the $\Delta\beta_p$ needed to produce the resultant 
$E_i^{(\rm final)}$.  We chose a few of the parameters to vary, and a few
essential features of the resonance structure, as given in the data 
or read off from Fig.~\ref{C12data}, in an attempt to produce a fit to that data.

This has turned out to be a rather complex task because of the interplay among 
the parameters, as well as due to some lack of flexibility of the collective 
model on which this analysis is based. It can be compared to squeezing a 
balloon \ldots \ something always pops out that one wants to keep in.  In fact,
we carried out this process in two stages. In the first, a quite rough agreement
with the data in Fig.~\ref{C12data} was found. This set of parameters was used 
as a ``new base'' and the process was repeated. In the end, we obtain a quite 
good description of the experimental cross section. But some small defects 
remain, so this result may not be the best that can be achieved.  With this 
multi-dimensional non-linear problem, there may be a number of ``quite good'' 
sets of parameters; we have found one.

\begin{figure}
\includegraphics[angle=0.0,scale=0.7]{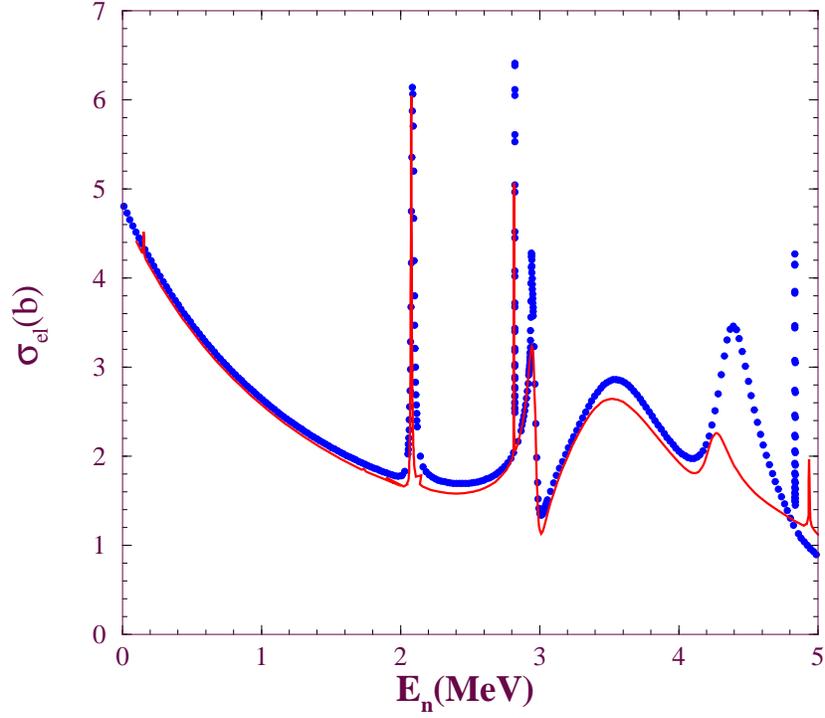}
\caption{\label{Theocross}
Comparisons between theory (dotted curve) and experimental data 
(solid curve, ENDF evaluated) of the 
cross section from the elastic scattering of 
neutrons from $^{12}$C.}
\end{figure}

Rather than attempting to fit everything, we decided to limit ourselves to 
trying to reproduce the most prominent features of the experimental elastic 
cross section shown in Fig.~\ref{C12data}, and of the experimentally best 
established resonances.  Specifically, 
we focused on the two ${3\over 2}^+$ resonances in the range 2.5 to 4.0 MeV 
energy, the prominent narrow ${5\over 2}^+$ resonance just above 2.0 MeV, and 
the known bound states below threshold.  The collective model used produces two 
${3\over 2}^+$ resonances in agreement with the data.
Of the two, one is very wide ($\Gamma \sim 1$ MeV), likely single-particle, and
can account for the broad peak centered around 3.5 MeV. The other is narrow, 
generally less than 100 keV, and by interference with the first, produces the 
prominent structure in the cross section near 3.0 MeV.  This is shown by the 
continuous curve in Fig.~\ref{Theocross}.  In this region a 
narrow ${1\over 2}^-$ resonance also is obtained from the model. Such is not 
seen in the data nor does a partner state exist in the spectrum of $^{13}$C. 
Note that in  Fig.~\ref{Theocross} we use the evaluated cross section
data file (ENDF) 
also found in the website of the National Nuclear Data Center,
Brookhaven ({\it www.nndc.bnl.gov}).
\begin{figure}
\includegraphics[angle=0.0,scale=0.7]{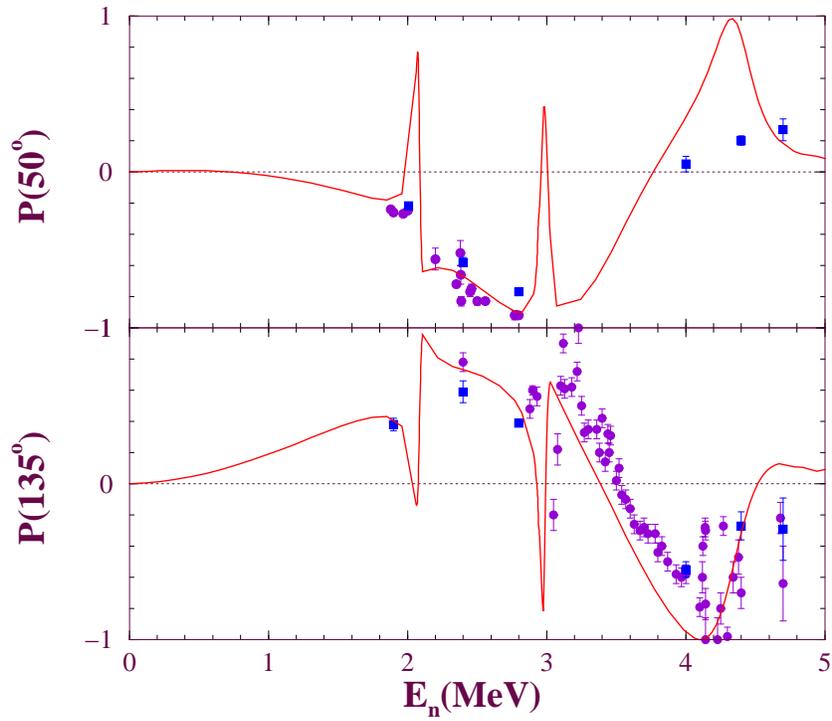}
\caption{\label{Theopol}
Comparisons between theory (solid curves) and experimental data 
(filled circles and squares with error bars) of the 
polarizations from the elastic scattering of 
neutrons from $^{12}$C.}
\end{figure}

Nevertheless, and as can be seen in Fig.~\ref{Theocross}, the overall agreement between experimental cross-section data and theory is good.  The theoretical
curve shows the structure just below 3.0 MeV as very similar in shape. The 
prominent narrow ${5\over 2}^+$ is reproduced at the correct energy. The smaller
${7\over 2}^+$ is observed in this figure too, and it is near the correct 
position.  At low energy above threshold, the
theoretical cross section is in good agreement with the smooth background
curve seen in the experimental data.  The theoretical curve also displays
the bump above 4.0 MeV, though it is higher and broader than the data.  This 
resonance is assessed to have spin-parity of ${1\over 2}^-$.  A more careful 
study of the negative-parity resonances at higher energy is required.  An 
appropriate treatment of higher-energy negative-parity states in the collective
model should include the $3^-$ state at 9.641 MeV in our selected set of 
channels. By so doing, the dimensionality of the problem increases and computing
time with the existing code becomes exorbitant. A new code based on massive 
parallel architecture is required. Currently such is under construction.

Polarization data (at two scattering angles) are shown in 
Fig.~\ref{Theopol}. They are compared with results that
come from the $S$ matrices determined by just seeking cross section features.  
The known resonances and a characteristic change of sign through the broad
$\frac{3}{2}^+$  resonance centered near 3.5 MeV are well replicated, as
is the variation attributable to that centered near 3.0 MeV. We stress that
these polarization data were not used in seeking a best  set of potential
parameters and were the result of but a single calculation (for each scattering
angle) using the cross section defined $S$-matrices.

The full set of resonance energies and widths and bound-state energies between
-35.0 MeV and +5.0 MeV obtained with these parameters are given in
Table~\ref{TabnewR}. It is worth stressing that only when Pauli blocking is 
treated by including the OPP does the approach give four states of the correct 
spin-parity and at appropriate energies below the n+$^{12}$C threshold.
Results shown in Table~\ref{TabnewR} 
for the spectra in the continuum are those obtained using the OPP terms
and are also those displayed in Fig.~\ref{Theocross}.  
\begin{table}
\begin{ruledtabular}
\caption{
\label{TabnewR}
Resonances and bound states found using the resonance identifier equations
and the parameter set of Eq.~(\ref{newParams}).}
\begin{tabular}{|c|cc|cc|cc|}
  $J^\pi$ & Expt. $E_R$~\cite{Aj91} & Expt.$\left({1\over 2}\Gamma_R\right)$
   & $E_R$ & $\left({1\over 2}\Gamma_R\right)$
   & $E_R$ & $\left({1\over 2}\Gamma_R\right)$ \\
\hline
    & Resonances & & with OPP & & without OPP\\
$({1\over 2})^+$ & 4.259 $\pm$ 0.015 & 0.11 & 4.833 &  0.53 & -- & \\
$({3\over 2})^+$ & 2.9 $\pm$ 0.01 4 & 0.062 & 2.965  & 0.030 & 2.997 & 0.025\\
$({3\over 2})^+$ & 3.472 $\pm$ 0.015 & 0.5  & 3.637  & 0.54 & 3.642 & 0.54 \\
$({5\over 2})^+$ & 2.079 $\pm$ 0.003 & 0.003 & 2.086  & 0.015 & 2.176 & 0.013\\
$({5\over 2})^+$ & -- & -- & 4.370  & 0.15 & 4.389 & 0.16 \\
$({7\over 2})^+$ & 2.77 & & 2.823  & 8.9 $\times 10^{-7}$ & 2.823 &8.9
$\times 10^{-7}$ \\
$({9\over 2})^+$ & 4.934 & 0.001 & 4.836  & 6.2 $\times 10^{-4}$
& 4.836  & 6.2 $\times 10^{-4}$ \\
$({1\over 2})^-$ & --  & -- & 2.939 &  5.0 $\times 10^{-3}$ & 0.0495
& 1.2 $\times 10^{-3}$ \\
$({3\over 2})^-$ & --  & -- & -- &  & 2.185  & 0.049\\
$({5\over 2})^-$ & $\simeq +0.1?$      & & --       & & 2.615
& 8.4 $\times 10^{-6}$ \\
\hline
                    & Bound states   & & with OPP& & without OPP &\\
   $({1\over 2})^-$ & -4.98          & & -4.8650 & &-5.9360  & \\
   $({1\over 2})^+$ & -2.0301        & & -1.9920 & &-1.8770  & \\
   $({1\over 2})^+$ & --             & & --       & &-18.9614  & \\
   $({1\over 2})^+$ & --             & & --       & &-26.5874 & \\
   $({3\over 2})^-$ & -1.3671        & & -1.4640 & &-0.9433 & \\
   $({3\over 2})^-$ & --             & & --       & &-3.6152 & \\
   $({3\over 2})^-$ & --             & & --       & &-10.0836 & \\
   $({3\over 2})^+$ & --             & & --       & &-21.9417 & \\
   $({5\over 2})^-$ & $\simeq +0.1$  & & -0.0195& &-7.9049& \\
   $({5\over 2})^+$ & -1.1833        & & -1.8720 & &-1.8178 & \\
   $({5\over 2})^+$ & --             & & --       & &-22.0019 & \\
   $({7\over 2})^-$ & --             & & --       & &-6.3119 & \\
   \end{tabular}
   \end{ruledtabular}
   \end{table}

The last two columns in Table~\ref{TabnewR} gives the results obtained
when Pauli blocking is ignored (by omitting the effects of the OPP).
Those results are identified also by the header, ``without OPP''. 
The bound state spectrum that results most clearly shows the spuriosity.
Far more states result than are known empirically (for $^{13}$C), 
and most are more deeply bound than the known ground state. With the
scattering cross sections, that spuriosity is not as well displayed.
 Indeed by adjusting parameter values we could attain nearly as good
a result as found when the OPP is used.  However, without the OPP
the resonance states are not the correct entries in the coupled-channel
sequence.  For example, without the OPP, the narrow $\frac{5}{2}$ resonance
near 2.0 MeV is the third in sequence rather than the second when Pauli
blocking is used. But, the high spin-parity resonances
should not be, and are not, affected by Pauli blocking.  For example,
in both calculated spectra the $\frac{9}{2}$ resonance has the same centroid
and width values.

Focusing on the results with the OPP corrections, clearly
the essential features of the even-parity resonances are reproduced well.
However, this analysis suggests that the experimental resonance at 4.259 MeV
would have a $J^\pi$ assignment of $5/2^+$ and that a broad resonance that
does not influence the elastic scattering has a centroid at 4.833 MeV
and would have a spin-parity of $1/2^+$.  Experiment suggests that
the $1/2^+$ is observed and lies at 4.259 MeV while the $5/2^+$ is not
to be seen in this energy regime.
But it must be remembered that we have limited the process to just three
states in ${}^{12}$C, and have some allowance in the parameter specifications 
in the collective model prescription for the interaction potentials.
Likewise there are two negative-parity resonant states indicated from
these calculations; namely the $1/2^-$ ``hidden'' in the structure above 
3.0 MeV and a $5/2^-$ which might align with what is a suggestion of a 
small narrow resonance just above threshold in the data.

Despite the few anomalies, it is clear that this
method makes  possible exploration of resonance behavior,
in this first presentation by using the collective 
model, and by using purely algebraic means. 
However, it is also clear that the collective model, coupled with 
the restrictions imposed by limiting the choice of active target
nucleus levels, will never provide as complete agreement with 
experimental data as one would like.  
It is planned to define interaction potential matrices from folding 
a suitable low energy $NN$ force with microscopic model structure 
such as given by the shell model; a process that has had much success
in recent years in analyses of higher energy nucleon-nucleus 
scattering~\cite{Am00}. Of course Pauli effects would be approximately
included via this folding when full antisymmetrization is considered.
Therefore, it is of interest at this stage to speculate what might
be learned from a
more microscopic theory, and that is discussed in the following section.

\section{The structure of ${}^{13}$C in terms of ${}^{12}$C plus a neutron}
\label{shellmodel}

The algebraic scattering program has been applied to low energy 
neutron scattering from ${}^{12}$C.  The energy variation of the measured 
cross section reveals sharp resonances, some of which one may associate 
with the presence of states in $^{13}$C.
Likewise, the algebraic methods allow prediction of bound states of 
the n-$^{12}$C system which should correspond to states in ${}^{13}$C 
below threshold.  In our first
calculation, the structure of $0^+_1$ (ground), $2^+_1$ (4.4389 MeV)  and 
$0^+_2$ (7.65 MeV) states of
$^{12}$C have been taken into account. That such a model should give results
characterizing the observed cross section follows from consideration 
of ($p$-$s$-$d$) shell models for $^{12,13}$C.

The key quantity that relates the structure of the spectrum of 
$^{13}$C to the spectrum
of $^{12}$C plus a neutron in a single particle orbit $j$ is the 
pickup spectroscopic amplitude,
\begin{equation}
S_{\lambda}^{jIJ} = \left\langle {\Psi_J^{(\lambda)}(1,2,\cdots, 13)} \left\|
{{\bf a}_j^\dagger} \right\| {\Phi_{I}(1,2,\cdots, 12)}\right\rangle .
\end{equation}

The development applies equally to a proton with the compound system then being
$^{13}$N.  The particle coordinates will be omitted hereafter with 
the identification:
\begin{eqnarray}
&&\vert \Psi_J^{(\lambda)} \rangle = {\rm the\ }\lambda^{\rm th} 
{\rm state\  of \    }
{}^{13}{\rm C\ with\ spin-parity\ } J^{\pi}
\nonumber\\
&&\vert \Phi_I \rangle = {\rm the\  state\ in\ } {}^{12}{\rm C\  with \ spin\ }
I\ .
\end{eqnarray}
Thus, while just three unique spin states in $^{12}$C are considered, 
there will be more
than 1 state of spin $J^{\pi}$ formed by attaching any specific orbit 
nucleon.  Whether
these numbers are predicted by a shell model calculation or instead 
are extracted from
scattering data analysis, there are two sum rules that indicate if 
the coupling of any
single particle state in the spectrum has been exhausted.

The first is the pickup sum rule. It is the result of summing the spectroscopic
probabilities over all target ($^{12}$C) states and is
\begin{eqnarray}
\Sigma_P = \sum_I \left( S_{\lambda}^{jIJ} \right)^2 &=&  \sum_I \left(
\left\langle {\Psi_J^{(\lambda)}} \left\| {{\bf a}_j^\dagger} 
\right\| {\Phi_{I}}
\right\rangle \right)^2
\nonumber\\
&=&  \sum_{IKm} \left[  (-)^{2j} \left\langle I K j m | J M\right\rangle
\ \left\langle {\Psi_J^{(\lambda)}} \left\| {\bf a}_j^\dagger 
\right\| {\Phi_{I}}
\right\rangle \right]^2\
\nonumber\\
&=& \sum_{IKm}\ (2J+1)^2 \ \left\langle \Psi_{JM}^{(\lambda)} \left| 
a_{jm}^\dagger
\right| \Phi_{IK} \right\rangle \ \left\langle \Phi_{IK} \left| a_{jm} \right|
\Psi_{JM}^{(\lambda)}\right\rangle .
\end{eqnarray}
On closure over the target states, this gives
\begin{equation}
\Sigma_P = (2J + 1)^2\ \sum_m \left\langle \Psi_{JM}^{(\lambda)} \left|
a_{jm}^\dagger a_{jm}\right| \Psi_{JM}^{(\lambda)} \right\rangle =
(2J+1)^2\  n^{(\lambda)}_j\ ,
\end{equation}
where $n^{(\lambda)}_j$ is the number of nucleons of the appropriate 
orbit in the
particular $^{13}$C state $\vert \Psi_{JM}^{(\lambda)} \rangle$.  A 
further summation over all possible values of $j$ gives
\begin{equation}
\sum_{jI}  \left( S_{\lambda}^{jIJ} \right)^2 = (2J+1)^2 N\ ,
\end{equation}
where N is the number of nucleons of the selected type (neutrons in the case
considered).
For the $^{13}$C-[$^{12}$C + n] systems, spectroscopic amplitudes have been
calculated~\cite{Ri80} using a shell model in which the active shells 
were $0p$-$1s$-$0d$.
Recently, the structure of $^{12}$C has been found~\cite{Ka95} using 
a complete no-core
$(0+2)\hbar \omega$ shell-model.  Such has not been used to specify 
$^{13}$C as yet.
However, the basic details should not be too different to what is 
discussed.  In $S_{(\lambda)}^{jIJ}$ expressing the
states of $^{13}$C in terms of $^{12}$C plus a neutron are given. 
They are presented in
the order of excitation in $^{13}$C with the excitation energy shown 
in column 2.
The equivalent energy in the center of mass for n + $^{12}$C 
is shown in column 3.  The first four entries therefore are closed to 
neutron scattering from $^{12}$C.
In the incident energy regime to the threshold of excitation of the 
2$^+_1$, (4.4389 MeV)
state in $^{12}$C, four known resonances are expected coinciding with 
the spin-parities
of the compound system being ${\frac {5}{2}}^+\vert_{(2)}; {\frac 
{3}{2}}^+\vert_{(1)};
{\frac {3}{2}}^+\vert_{(2)}$ and ${\frac {1}{2}}^-\vert_{(2)}$.  The 
third of those however is very broad, $\cal O$(MeV), and $R$-matrix 
studies~\cite{Ri80} suggest that it is a single particle potential resonance.

Thus the shell model indicates that the four known states in $^{13}$C
below the n+$^{12}$C threshold are well represented by a neutron coupled to the
ground and $2^+_1$ states of ${}^{12}$C and so should be (and are) well defined
by the negative energy solutions of the multichannel algebraic scattering
problem. 
Likewise, coupling a neutron to the same states in $^{12}$C give shell model
candidates for states in $^{13}$C above the n+${}^{12}$C threshold; ones that match 
observations from scattering data.  They relate to resonances in the scattering
of neutrons from ${}^{12}$C which the multichannel scattering theory also
defines well.  The sizeable components of a neutron coupled to the $2^+$ state
of $^{12}$C that the shell model indicates is consistent with the strong
coupling we have found necessary via a collective model prescription in
these evaluations of scattering observables.

The second sum rule, the stripping sum rule, is formed by completing 
a sum over all compound mass ($^{13}$C) states.
It is important as it indicates whether or not the model calculation
may give the large components of any state.  That sum rule is defined by
\begin{eqnarray}
\Sigma_S = \sum_{J, \lambda} \left( S_{(\lambda)}^{jIJ} \right)^2 &=&
\sum_{J, M;\lambda m} \left[  (-)^{2j} \left\langle J M j m | I K \right\rangle
\left\langle \Psi_J^{(\lambda)} \left\|\bf a_j^\dagger \right\| 
\Phi_{I} \right\rangle
  \right]^2\ ,
\nonumber\\
&=&
\sum_{J,M;\lambda m} (2I+1)^2
\left\langle \Phi_{IK} \left| a_{jm} \right| \Psi_{JM}^{(\lambda)} 
\right\rangle
\ \left\langle \Psi_{JM}^{(\lambda)} \left| a_{jm}^\dagger \right| \Phi_{IK}
\right\rangle
\nonumber\\
&=&
(2I+1)^2 \sum_m
\left\langle \phi_{IK} \left| a_{jm} a_{jm}^\dagger \right| 
\Phi_{IK}\right\rangle\ .
\end{eqnarray}
Using the (anti)commutation property of the creation/annihilation 
operators this sum
reduces to
\begin{equation}
\Sigma_S = (2I+1)^2 \left[ 2j + 1 - n_j \right]\ ,
\end{equation}
where  $n_j$ is the number of nucleons in the orbital $j$ in the 
target ($^{12}$C) state $\vert \Phi_{IK}\rangle$.  In practical calculations 
one cannot deal with all of the mass 13 levels and so this sum rule is best 
viewed as an  estimate of how much
transition strength lies with states other than those investigated.

Using the $p$-$s$-$d$ shell model gives
stripping sum rule values listed in Table~\ref{Striptab}. 
They are compared against a theoretical limit set assuming that there 
are no $1s-0d$ neutrons in the ground state description of ${}^{12}$C.  
That is quite reasonable as a shell model
study~\cite{Ka95} made using the complete $(0+2)\hbar \omega$ space 
gives 11.6 nucleons within the $0s-0p$ shells for both the $0^+(gs)$ and 
$2^+$(4.4389) states.  However, with the simplest of shell models 
(packed orbits), the pickup sum rule values theoretically
are 2, 4, and 6 for the $1s_{1/2}, 0d_{3/2}$, and $0d_{5/2}$ shells 
respectively.  The occupancies of the $0s_{1/2}, 0p_{3/2}, 0p_{1/2}$ used 
to get the results listed in Table~\ref{Striptab} are those  given by 
$p$-$s$-$d$ shell model calculations~\cite{Ri80},
namely 2.0, 3.288 and 0.712 respectively.
\begin{table}
\caption{\label{Spectab} Shell model values for pickup spectroscopic 
amplitudes giving $^{13}$C
for neutrons on ${}^{12}$C with energies to 4.0 MeV.  Single particle 
states belong to the
$0p-0d-1s$ space.}
\begin{ruledtabular}
\begin{tabular}{|c|c|c|cc|cc|}
\hspace*{0.2cm}${}^{13}$C\hspace*{0.2cm} &
\hspace*{0.2cm} E$_{ex}$\hspace*{0.2cm} &
\hspace*{0.2cm} E rel. C.M.\hspace*{0.2cm} &
\hspace*{0.1cm}neutron\hspace*{0.1cm} &
\hspace*{0.1cm} ${}^{12}$C\hspace*{0.1cm} &
\hspace*{0.1cm}One channel\hspace*{0.1cm} &
\hspace*{0.1cm} Two channels\hspace*{0.1cm} \\
${J}^\pi_{(\lambda)}$ &
in ${}^{13}$C &
n + ${}^{12}$C &
$\ell_j$ &
  ${I}$ &
$S_{(\lambda)}^{j0j}$ &
$S_{(\lambda)}^{jIJ}$ \\
\hline
$\frac {1}{2}^-\vert_1$ &\ 0.00 & -4.95 & $p_{1/2}$ & 0 &\ 1.127 &\ 1.107 \\
  & & & $p_{3/2}$ & 2 & & -1.498 \\
$\frac {1}{2}^+\vert_1$ &\ 3.09 & -1.86 & $s_{1/2}$ & 0 &\ 1.349 &\ 1.349 \\
  & & & $d_{3/2}$ & 2 & & -0.148 \\
  & & & $d_{5/2}$ & 2 & &\ 0.069 \\
$\frac {3}{2}^-\vert_1$ &\ 3.68 & -1.26 & $p_{3/2}$ & 0 & -0.840 & -0.887 \\
  & & & $p_{1/2}$ & 2 & & -1.819 \\
  & & & $p_{3/2}$ & 2 & &\ 1.122 \\
$\frac {5}{2}^+\vert_1$ &\ 3.85 & -1.09 & $d_{5/2}$ & 0 &\ 2.271  &\ 2.271 \\
  & & & $s_{1/2}$ & 2 & &\ 0.334 \\
  & & & $d_{3/2}$ & 2 & & -0.293 \\
  & & & $d_{5/2}$ & 2 & &\ 0.415 \\
$\frac {5}{2}^+\vert_2$ &\ 6.86 &\ 1.92 & $d_{5/2}$ & 0 &\ 0.409  &\ 0.409 \\
  & & & $s_{1/2}$ & 2 & & -2.173 \\
  & & & $d_{3/2}$ & 2 & &\ 0.234 \\
  & & & $d_{5/2}$ & 2 & &\ 0.502 \\
$\frac {7}{2}^+\vert_1$ &\ 7.49 &\ 2.54 & $d_{3/2}$ & 2 & &\ 0.778 \\
  & & & $d_{5/2}$ & 2 & & -2.565 \\
$\frac {5}{2}^-\vert_1$ &\ 7.55 &\ 2.60 & $p_{1/2}$ & 2 & &\ 1.314 \\
  & & & $p_{3/2}$ & 2 & &\ 0.495 \\
$\frac {3}{2}^+\vert_1$ &\ 7.68 &\ 2.73 & $d_{3/2}$ & 0 &\ 0.944 &\ 0.944 \\
  & & & $s_{1/2}$ & 2 & &\ 1.638 \\
  & & & $d_{3/2}$ & 2 & &\ 0.192 \\
  & & & $d_{5/2}$ & 2 & &\ 0.326 \\
$\frac {3}{2}^+\vert_2$ &\ 8.20 &\ 3.30 & $d_{3/2}$ & 0 & -1.000 & -1.000 \\
  & & & $s_{1/2}$ & 2 & &\ 0.803 \\
  & & & $d_{3/2}$ & 2 & &\ 0.318 \\
  & & & $d_{5/2}$ & 2 & & -0.644\\
$\frac {1}{2}^-\vert_2$ &\ 8.86 &\ 3.91 & $p_{1/2}$ & 0 &\ 0.082 & -0.049 \\
  & & & $p_{3/2}$ & 2 & &\ 0.084 \\
$\frac {3}{2}^-\vert_2$ &\ 9.50 &\ 4.55 & $p_{3/2}$ & 0 &\ 0.052  & -0.075 \\
  & & & $p_{1/2}$ & 2 & & -0.102 \\
  & & & $p_{3/2}$ & 2 & &\ 1.123 \\
$\frac {3}{2}^-\vert_3$ &\ 9.90 &\ 4.95 & $p_{3/2}$ & 0 & &\ 0.031 \\
  & & & $p_{1/2}$ & 2 & &\ 0.058 \\
  & & & $p_{3/2}$ & 2 & & -0.205 \\
\end{tabular}
\end{ruledtabular}
\end{table}
For those single particle specifics the theoretical sum rule limits 
are 0.0, 0.712, and 1.288.  In the $2^+$ (4.4389 MeV) state in ${}^{12}$C, 
the occupancies are varied slightly from those with the $0p$ shell numbers 
most altered.  The relevant theoretical sum rules then were found using 
occupancies of 2.0, 3.01, and 0.98.  The sum rule exhaustion
suggests that the assumption that the states in the listings above 
are a nucleon plus the ${}^{12}$C nucleus in either the ground or the 
2$^+$ (4.4389 MeV) state is quite well satisfied except if that nucleon 
is in the $d_{\frac{3}{2}}$ orbit. It would not surprise, therefore, if
the excitation of the ${}^{13}$C $\frac{3}{2}^+$ states may
not be as well described as others with any model involving the 
ground and 2$^+$ states of ${}^{12}$C in the basis.  Note also from 
Table~\ref{Spectab} that the $\frac{7}{2}^+$ and $\frac{5}{2}^-$ states 
anticipated to lie at 2.54 and 2.6 MeV excitation in the
n-${}^{12}$C system, are solely based upon couplings with the 
${}^{12}$C 2$^+$ state.  As such they may be only weakly excited in the 
n-${}^{12}$C scattering cross section.
\begin{table}
\caption{\label{Striptab}
The sum rules for the two channel case -- neutron space $0p-1s-0d$.}
\begin{ruledtabular}
\begin{tabular}{|c|ccc|c|}
\hspace*{0.2cm}Neutron\hspace*{0.2cm} &
\hspace*{0.2cm} ${}^{12}$C\hspace*{0.2cm} &
\hspace*{0.1cm} Theoretical\hspace*{0.1cm} &
\hspace*{0.1cm} Shell model\hspace*{0.1cm} &
\hspace*{0.1cm} exhaustion\hspace*{0.1cm} \\
state & state & sum rule & sum rule & \% \\
\hline
$s_{1/2}$ & $0^+$ & 2 & 1.820 & 91 \\
           & $2^+$ & 10 & 8.387 & 84 \\
$p_{1/2}$ & $0^+$ & 1.240 & 1.239  & 100 \\
           & $2^+$ & 5.080 & 5.050 & 99 \\
$p_{3/2}$ & $0^+$ & 0.763 & 0.758 & 99 \\
           & $2^+$ & 4.935 & 4.871 & 99 \\
$d_{3/2}$ & $0^+$ & 4 & 3.110 & 78 \\
           & $2^+$ & 20 & 0.95 &  5 \\
$d_{5/2}$ & $0^+$ & 6 & 5.557 & 93 \\
           & $2^+$ & 30 & 25.281 & 84 \\
\hline
\end{tabular}
\end{ruledtabular}
\end{table}

\section{conclusions}

A model for nucleon-nucleus scattering that uses sturmian expansions of 
multichannel interactions between the colliding nuclei has been used and 
found to give resonance scattering upon an average background; typical of 
what is measured with low-energy experiments.  That sturmian expansion 
theory also gives resonance identifier equations whose solutions identify 
feasibly, all possible resonance aspects to the scattering problem. Indeed 
we have found that it is quite practical to find the narrow resonances
(compound and quasi-compound) of the coupled-channel Schr\"{o}dinger problem 
by diagonalizing the energy-dependent matrix $ \mbox{\boldmath 
$\eta$}^{-{1\over 2}} {\bf G}_0(E) \mbox{\boldmath $\eta$}^{-{1\over 2}}$ 
and studying the trajectories of the relevant eigenvalues in the Gauss plane. 
When one of these complex eigenvalues evolves past the point $1+i0$ and does 
so having also a small imaginary component, one of these resonances occurs. 
Then Eqs.~(\ref{e-res}) and (ref{g-res}) can be used to determine
the resonance parameters.  Since these eigenvalues have smooth 
energy dependencies, it is generally much simpler so to determine the 
occurrence of resonance states than by other means that have been used to date.

We have used a collective model (to second order) to define a multichannel 
potential matrix for low-energy neutron-$^{12}$C scattering allowing coupling 
between the $0^+_1$ (ground), $2^+_1$ (4.4389 MeV), and $0^+_2$ (7.64 MeV) 
states with coupling taken to second order. The algebraic $S$ matrix for this 
system has been evaluated.  Good results have been found for the sub-threshold
bound states and for the  cross sections and polarizations  as functions
of energy.  The latter are rich in structure having both narrow and broad 
resonant features for different $J^\pi$; the existence and parameter values 
of which can be ascertained with ease.

The introduction of the Pauli exclusion principle in the coupled-channel
model by means of the orthogonalizing pseudo-potential (OPP) technique
ensures that the  model gives  a spectrum (bound states and resonances) 
that are built as physical states.  This was an important improvement giving 
an unique identification of states with respect to experimental data.

The results indicate that this approach has predictive power and can be used 
to interpret the experimental resonance spectra of the nuclear processes at 
low energy.  The discussion on the n-$^{12}$C resonant spectrum contained herein
is just one simple example of the potential use of this approach in nuclear 
analyses.


\begin{acknowledgments}
This research was supported by a grant from the Australian Research Council, 
by a merit award with the Australian Partners for Advanced Computing, by the 
Italian MURST-PRIN Project ``Fisica Teorica del Nucleo e dei Sistemi a Pi\`u 
Corpi'', and by the Natural Sciences and Engineering Research Council (NSERC), 
Canada. JPS, KA, and DvdK acknowledge the hospitality and support of the INFN, 
Padova, and of the Dipartimento di Fisica, Universit\`a di Padova between 1999 
and 2003.  LC and GP also would like to thank the School of Physics, University of 
Melbourne, and the Department of Physics and Astronomy, University of Manitoba
for hospitality and support.  The authors also thank Dr. P. J. Dortmans for 
constructing the base version of the codes we have used to evaluate 
multi-channel algebraic scattering matrices. 
\end{acknowledgments}


\appendix

\section{The first generation sturmians}
\label{stur-gen}

With $l$ being the orbital angular momentum quantum number that is 
embedded in the set collectively denoted by $c$, the eigenvalues $q_{li}$ 
of the first generation sturmian problem may be  calculated  as follows.
Consider a square well problem for a binding energy $B$ and well with 
depth ${\cal V}_0$ and radius $R$.
With $x_{li} = q_{li} \rho, {\bar q} = \sqrt{\mu B}$, and 
$y = {\bar q} \rho$, the first eigenvalues are defined from solution of 
the transcendental equations,
\begin{equation}
x_{li} \cot x_{li}+y=0\ ,
\end{equation}
and
\begin{equation}
x_{li}{F_{l-1}(x_{ln}) \over F_l(x_{li})} +
y{\Omega_{l-1}(y) \over \Omega_l(y)} = 0 \ ; \ l>0\ .
\end{equation}
The functions $F_l$ and $\Omega_l$ are the Riccati--Bessel and 
modified Riccati-Bessel (of the 3$^{rd}$ kind) functions respectively; 
each of order $l$.  They can be solved using well known recursion 
formulas~\cite{Pi86}. In terms of these, the first generation sturmians are
\begin{equation}
\Phi_{li}^0(r) = A_{li}
\left\{
\begin{array}{cr}
F_l(q_{ln}r) & \hspace*{0.5cm}{\rm for}\  0 \leq r \leq \rho \\
\left[\frac{F_l(q_{li}\rho)}{\Omega_l(y)}\right]
\Omega_l(\bar q r) & \hspace*{0.5cm}{\rm for}\  r \geq \rho\ .
\end{array}
\right. 
\end{equation}
The normalization constant is
\begin{equation}
A_{li}=\sqrt{\frac{\mu}{(q_{li}^2+\bar q^2) \int_0^\rho F_l^2(q_{li}r)dr}}\ ,
\end{equation}
and the sturmian eigenvalue is
\begin{equation}
\eta^{(0)}_{ci} = \frac{\mu {\cal V}_0}{q^2_{li} + {\bar q}^2 }\ .
\end{equation}
Note that the well depth ${\cal V}_0$ plays no role in the development
of the separable expansions as the product $U_c^0 
\left[\eta^{(0)}_{ci} \right]^{-1}$ assures that it does not carry through.

\section{The optical potential}
\label{Append-one}

Assuming a local form for the elastic channel element of the 
potential matrix, the
optical potential for elastic scattering is defined by
\begin{eqnarray}
V^{opt}(r,r';E) &=& V_{1 1}(r) + \sum^C_{c,c'=2} V_{1 c}(r)\ 
G^{(Q)}_{cc'}(r,r';E)\
V_{c' 1}(r')
\nonumber\\
&=& V_{1 1}(r) + \Delta U(r,r';E)\ .
\label{opteq}
\end{eqnarray}
Here $\Delta U$, the dynamic polarization potential (DPP), makes this 
formulated
interaction complex, nonlocal, and energy dependent as 
$G^{(Q)}_{cc'}$ are the full
Green functions referring to the $Q\ (= C -1)$ excluded channels. 
Those Green functions
are solutions of LS type equations built upon the free single channel 
Green function,
$G_{0c}$, namely
\begin{equation}
G^{(Q)}_{cc'} = G_{0c} \delta_{cc'} + \sum^C_{c'' = 2}\ G_{0c} \ V_{cc''}\ 
G^{(Q)}_{c''c'}\ .
\label{LStype}
\end{equation}
This complex equation is vastly simplified when the interactions are 
approximated by
separable expansions of finite rank and then~\cite{Ca87,Ca88,Ca90}
\begin{equation}
\Delta U(r,r';E) = \sum^N_{n, n' = 1} \chi_{1 n}(r) 
\left[{\mbox{\boldmath $\Lambda$}}(E) \right]_{nn'}(E) \chi_{n' 1}(r')\ ,
\label{DPP}
\end{equation}
where
\begin{equation}
{\mbox{\boldmath $\Lambda$}}(E) = 
\left[ {\mbox{\boldmath $\eta$}} - {\bf G}_0^{(Q)}(E)\right]^{-1} 
- {\mbox{\boldmath $\eta$}}^{-1}
\label{green_r}
\end{equation}
involves
\begin{equation}
\left[{\bf  G}_0^{(Q)}(E)\right]_{n n'} = 
\mu \left[ \sum^{\rm open}_{c\ne 1} \int_0^\infty
\frac {{\hat \chi}_{cn}(x) {\hat \chi}_{cn'}(x)} 
{k_c^2 - x^2 + i\epsilon} x^2 dx
- \sum^{\rm closed}_{c\ne 1} \int_0^\infty 
\frac {{\hat \chi}_{cn}(x)
{\hat \chi}_{cn'}(x)} 
{h_c^2 + x^2} x^2 dx \right]\ ,
\label{curlyG}
\end{equation}
where $k_c$ and $h_c$ are respectively the wave numbers relevant for 
each open and
closed channel considered.  In application~\cite{Ca91,Pi95}, various
perturbative-iterative schemes have been used to evaluate the DPP and thence to
determine the (elastic) scattering $S$ matrix.

\section{Local potential matrix elements}
\label{Append-two}

Using the Woods-Saxon forms
\begin{equation}
f(r) = \left[1 + e^{\left( {{r-R}\over a} \right)} \right]^{-1}
\hspace*{0.3cm} ; \hspace*{0.3cm} g(r) = \frac{1}{r} \frac{df(r)}{dr}
\hspace*{0.3cm} ; \hspace*{0.3cm} W_{ls} = 2 V_{ls} \lambda_\pi^2 \ ,
\end{equation}
with $\kappa_l = \beta_L R_0/a$ and the definitions
\begin{eqnarray}
e(r) &=& \exp{\left(\frac{r - R_0}{a}\right)} \hspace*{0.5cm};\hspace*{0.5cm}
f_0(r) = \left[ 1 + e(r) \right]^{-1} \nonumber\\
A(r) &=& e(r)\ \left[1 + e(r)\right]^{-2} = \frac{a}{R_0}
\left.\frac{df(r)}{d\epsilon} \right|_0 = -a r\ g_0(r)
\nonumber\\
B(r) &=& e(r)\ [e(r) -1]\  \left[1 + e(r)\right]^{-3}
= \left(\frac{a}{R_0}\right)^2\left.\frac{d^2f(r)}{d\epsilon^2} \right|_0
= - \left({a^2 r\over R_0}\right) \left.{dg\over d\epsilon}\right|_0
\nonumber\\
C(r) &=& \frac{e(r)}{\left[1 + e(r)\right]^4} \left[e^2(r) - 4e(r) +1 \right]
=-ar \left({a \over R_0}\right)^2 \left.{d^2g\over d\epsilon^2}\right|_0
\ ,
\end{eqnarray}
the potential matrix elements, Eq.~(\ref{potchan2}),  take the form, 
omitting the relevant parity indices,
\begin{eqnarray}
V_{c'c}(r) &=&
\left\{[V_0 + l(l+1)\ V_{ll}] \left[f_0(r) + 1/(8\pi)\ \kappa_L^2 
B(r)\right] \right.
\nonumber\\
&&\hspace*{3.0cm}- \frac{1}{ar}
\left. W_{ls} \left[A(r) + 1/(8\pi)\ 
\kappa_L^2 C(r)
\right] [{\bf \ell} {\bf  \cdot} {\bf s}]_c \right\}\delta_{c'c}
\nonumber\\
&&+ V_{ss} \left[ f_0(r) + 1/(8\pi)\ \kappa^2_L B(r)\right] \ [{\bf 
s} {\bf \cdot}
{\bf I}]_{c'c}
\nonumber \\
  && + \left\{V_0 + \frac{1}{2} V_{ll} [l'(l'+1) + l(l+1)]\right\}\
  \nonumber\\
   &&\hspace*{1.0cm} \times \left\{ [4\pi/(2L+1)]^{\frac{1}{2}}\ \kappa_L A(r)\
   \left[{\bf Y}_L {\bf \cdot} {\bf Y}_L\right]_{c'c} +
   \frac{1}{2} (2L + 1)^2 \kappa^2_L B(r)\right.
   \nonumber\\
   &&\hspace*{2.0cm} \left. \times
   \sum_{\ell = 2} \frac{1}{(2\ell + 1)} |\langle L 0 L 0 |\ell 0\rangle |^2
   \left[{\bf Y}_{\ell} {\bf \cdot} {\bf Y}_{\ell}\right]_{c'c} \right\}
   \nonumber \\
   &&- \frac {1}{2ar} W_{ls}  \left\{ [{\bf \ell} {\bf \cdot} {\bf s}]_{c'}
   + [{\bf \ell} {\bf \cdot} {\bf s}]_c \right\}
   \nonumber\\
   &&\hspace*{1.0cm} \times \left\{ [4\pi/(2L+1)]^{\frac{1}{2}}\ \kappa_L
   B(r) \ \left[{\bf Y}_L {\bf \cdot} {\bf Y}_L\right]_{c'c}
   + \frac{(2L+1)^2}{2} \kappa^2_l C(r)\right.\nonumber\\
   &&\hspace*{2.0cm} \left. \times
   \sum_{\ell = 2}
   \frac{1}{(2\ell + 1)} |\langle L 0 L 0 |\ell 0\rangle |^2\
   \left[{\bf Y}_{\ell} {\bf \cdot} {\bf Y}_{\ell}\right]_{c'c} \right\}
   \nonumber\\ 
  &&+ \frac{1}{2} V_{ss} [4\pi/(2L+1)]^{\frac{1}{2}}\ \kappa_L A(r)\
  \nonumber\\
  &&\hspace*{1.0cm}
  \times
  \sum_{c''} \left\{ [{\bf s} {\bf \cdot} {\bf I}]_{c'c''}
  \ \left[{\bf Y}_L {\bf \cdot} {\bf Y}_L\right]_{c''c} +
  \ \left[{\bf Y}_L {\bf \cdot} {\bf Y}_L\right]_{c'c''}
  [{\bf s} {\bf \cdot} {\bf I}]_{c''c} \right\}
  \nonumber\\
  &&+ \frac{1}{2} V_{ss} \frac{(2L+1)}{2} \kappa^2_L B(r)
  \left( \sum_{\ell = 2}
  \frac{1}{(2\ell + 1)} |\langle L 0 L 0 |\ell 0\rangle |^2\ \right. \nonumber\\
  && \hspace*{1.0cm}
  \times \left.
  \sum_{c''} \left\{ [{\bf s} {\bf \cdot} {\bf I}]_{c'c''}
  \left[{\bf Y}_{\ell} {\bf \cdot} {\bf Y}_{\ell}
  \right]_{c''c} + \left[{\bf Y}_{\ell} {\bf \cdot} {\bf Y}_{\ell}
  \right]_{c'c''}   [{\bf s} {\bf \cdot} {\bf I}]_{c''c} \right\}\right) \ .
  \label{vv}
  \end{eqnarray}

For the specific case considered, that of neutron scattering from ${}^{12}$C
and allowing coupling with the 0$^+$ (ground state), the $2^+$ 
(4.4389MeV), and the
excited  0$^+$ (7.64 MeV) states of ${}^{12}$C, the L = 2 multipole only
is needed in the deformation expansions. Using the specific values for
the parity Clebsch-Gordan coefficients of
\begin{equation}
\langle 2 0 2 0 | 0 0 \rangle = \frac{1}{\sqrt{5}}\ \ ;\ \
\langle 2 0 2 0 | 2 0 \rangle = -\sqrt{\frac{2}{7}}\ \ ;\ \
\langle 2 0 2 0 | 4 0 \rangle = 3 \sqrt{\frac{2}{35}}
\end{equation}
gives the result
\begin{eqnarray}
V_{c'c}(r) &=&
\left\{[V_0 + l(l+1)\ V_{ll}] \left[f_0(r) + 1/(8\pi)\ \kappa_2^2 
B(r)\right] \right.
\nonumber\\
&&\hspace*{3.0cm}- \frac {1}{ar}\left. W_{ls} \left[A(r) + 1/(8\pi)\ 
\kappa_2^2 C(r)
\right] [{\bf \ell} {\bf  \cdot} {\bf s}]_c \right\}\delta_{c'c}
\nonumber\\
&&+ V_{ss} \left[ f_0(r) + 1/(8\pi)\ \kappa^2_2 B(r)\right] \ [{\bf 
s} {\bf \cdot}
{\bf I}]_{c'c}
\nonumber\\
&&+ \left\{V_0 + \frac{1}{2} V_{ll} [l'(l'+1) + l(l+1)]\right\}\
\nonumber\\
&&\hspace*{0.5cm}\times \left\{
\left(\sqrt{(4\pi/5)} \kappa_2 A(r)
  + \frac{\kappa_2^2}{7} B(r)\right)
  \left[{\bf Y}_2 {\bf \cdot} {\bf Y}_2\right]_{c'c}
  + \frac{\kappa_2^2}{7} B(r) \left[{\bf Y}_4 {\bf \cdot} {\bf Y}_4\right]_{c'c}
  \right\}
  \nonumber\\
  &&- \frac {1}{2ar} W_{ls}  \left\{ [{\bf \ell} {\bf \cdot} {\bf s}]_{c'}
  + [{\bf \ell} {\bf \cdot} {\bf s}]_c \right\}
  \nonumber\\
  &&\hspace*{0.5cm} \times \left\{
  \left(\sqrt{(4\pi/5)} \kappa_2 B(r)
   + \frac{\kappa_2^2}{7} C(r)\right)
   \left[{\bf Y}_2 {\bf \cdot} {\bf Y}_2\right]_{c'c}
   + \frac{\kappa_2^2}{7} C(r) \left[{\bf Y}_4 {\bf \cdot} {\bf 
Y}_4\right]_{c'c}
   \right\}
   \nonumber\\                                        
   &&+
   \frac{1}{2}V_{ss} \left\{ \left( \sqrt{(4\pi/5)} \kappa_2 A(r)
   + \frac{\kappa_2^2}{7} B(r) \right) \right.
   \nonumber\\
   &&\hspace*{0.5cm} \times
   \sum_{c''} \left[ [{\bf s} {\bf \cdot} {\bf I}]_{c'c''}
   \ \left[{\bf Y}_2 {\bf \cdot} {\bf Y}_2\right]_{c''c} +
   \ \left[{\bf Y}_2 {\bf \cdot} {\bf Y}_2\right]_{c'c''}
   [{\bf s} {\bf \cdot} {\bf I}]_{c''c} \right]
   \nonumber\\
   &&\hspace*{1.0cm} \left.
    + \frac{\kappa_2^2}{7} B(r)
    \sum_{c''} \left[ [{\bf s} {\bf \cdot} {\bf I}]_{c'c''}
    \ \left[{\bf Y}_4 {\bf \cdot} {\bf Y}_4\right]_{c''c} +
    \ \left[{\bf Y}_4 {\bf \cdot} {\bf Y}_4\right]_{c'c''}
    [{\bf s} {\bf \cdot} {\bf I}]_{c''c} \right]
    \right\}  \, .
\label{fullV}
    \end{eqnarray}
 
In deriving Eq.~(\ref{fullV}) from Eq.~(\ref{potchan2}), use has been made
of the matrix elements:
\begin{equation}
\langle \mbox{\boldmath $\ell \cdot \ell$} \rangle =
\langle \ell' j' I' J | \mbox{\boldmath $\ell \cdot \ell$}
| \ell j I J \rangle
= \delta_{\ell\ell'}\delta_{jj'}\delta_{II'} \ell(\ell + 1)\ ,
\end{equation}
and
\begin{equation}
\langle \mbox{\boldmath $s \cdot \ell$} \rangle =
\delta_{\ell\ell'}\delta_{jj'}\delta_{II'} \times
\left\{
\begin{array}{ccc}
\frac{\ell}{2} & {\rm , if} & j=\ell+\frac{1}{2}\\
-\frac{\ell+1}{2} & {\rm , if} & j=\ell-\frac{1}{2}\\
\end{array}
\right.
.
\end{equation}
The spin-spin matrix element is a little more complicated~\cite{Va88},
namely:

\begin{eqnarray}
\langle {\bf s} {\bf \cdot} {\bf I} \rangle
&=& (-)^{(j+j'+J)} \left\{
\begin{array}{ccc}
j' & j & 1\\
I & I' & J\\
\end{array}
\right\}
\left\langle I' \left\| {\bf I} \right\| I \right\rangle
\left\langle s \left\| {\bf s} \right\| s \right\rangle
\nonumber \\
&=& \delta_{II'} \delta_{ll'}  (-)^{\left(1/2 + j - j'+ I + J + l\right)}
\sqrt{(2j+1) (2j'+1) (2I+1)} \nonumber\\
&&\hspace*{1.0cm}\times \sqrt{\frac{3}{2} I(I+1)}
\left\{
\begin{array}{ccc}
j' & j & 1\\
I & I & J\\
\end{array}
\right\}
\left\{
\begin{array}{ccc}
\frac{1}{2} & l & j\\
j' & 1 & \frac{1}{2}\\
\end{array}
\right\} \ .
\end{eqnarray}

Since the operator is diagonal in I and $\ell$,
and null if either I or I' is null, for our calculations
the following particular case is useful:

\begin{equation}
\langle \ell j' 2 J | {\bf s} {\bf \cdot} {\bf I}
| \ell j 2 J \rangle
= \frac{1}{2(2\ell+1)}K(\ell,j,j',J).
\end{equation}
The values of interest for $K$ are the following:
\begin{eqnarray}
K(\ell,\ell+\frac{1}{2},\ell+\frac{1}{2},J)&=&-\ell(\ell+2)+J(J+1)
-\frac{27}{4},\\
K(\ell,\ell-\frac{1}{2},\ell-\frac{1}{2},J)&=&\ell^2-J(J+1)
+\frac{23}{4},\\
K(\ell,\ell+\frac{1}{2},\ell-\frac{1}{2},J)&=&
K(\ell,\ell-\frac{1}{2},\ell+\frac{1}{2},J)
\\
&=&\sqrt{(J+\ell+\frac{7}{2})(J+\ell-\frac{3}{2})
(J-\ell+\frac{5}{2})(-J+\ell+\frac{5}{2})} \ .
\nonumber
\end{eqnarray}
We give, finally, the
matrix elements of the scalar product of two rank L spherical harmonics as
\begin{eqnarray}
\langle {\bf Y}_L {\bf \cdot} {\bf Y}_L\rangle &=&
\left\langle {l'j'I'J} \left| {\bf Y}_L(\hat r) {\bf \cdot}
{\bf Y}_L(\hat \Upsilon)
\right| ljIJ \right\rangle \nonumber\\
&=& (-)^{\left( j + I' + J \right)}
\left\{
\begin{array}{ccc}
j' & j & L\\
I & I' & J\\
\end{array}
\right\}
\left\langle \left(l'{\frac{1}{2}}\right)j' \right\| {\bf Y}_L(\hat r) \left\|
\left(l{\frac{1}{2}}\right)j \right\rangle
\left\langle I' \left\| {\bf Y}_L(\hat \Upsilon)
\right\| I \right\rangle \nonumber \\
&=& (-)^{\left(j+I'+l'-{\frac{1}{2}}\right)} \sqrt{(2j+1) (2j'+1) 
(2I+1) (2l+1)}
\frac {1}{4\pi} (2L+1) \nonumber\\
&& \hspace*{0.5cm}\times
\left\langle I 0 L 0 \vert I^\prime 0 \right\rangle
\left\langle l 0 L 0 \vert l^\prime 0 \right\rangle
\left\{
\begin{array}{ccc}
j' & j & L\\
I & I & J\\
\end{array}
\right\}
\left\{
\begin{array}{ccc}
l & \frac{1}{2} & j\\
j' & L & l'\\
\end{array}
\right\} \ ,
\label{tensor}
\end{eqnarray}
then, on using the identity
\begin{equation}
\left\{
\begin{array}{ccc}
l & \frac{1}{2} & j\\
j' & L & l'\\
\end{array}
\right\}
\left\langle l 0 L 0 \vert l^\prime 0 \right\rangle
= (-)^{\left( l+j'+\frac{1}{2}\right)} \frac {1}{\sqrt{(2l+1)(2j'+1)}}
\left\langle j {\frac{1}{2}} L 0 \right|\left. j^\prime {\frac{1}{2}} 
\right\rangle \ ,
\end{equation}
Eq.~(\ref{tensor}) reduces to
\begin{eqnarray}
\left\langle l'j'I'J \left| {\bf Y}_L({\hat r}) {\bf \cdot}
{\bf Y}_L({\hat \Upsilon}) \right|
ljIJ \right\rangle &=& (-)^{\left(J-{\frac{1}{2}}\right)} \frac {1}{4\pi}
\sqrt{(2I+1) (2j+1) (2j'+1) (2L+1)} \nonumber\\
&&\times
\left\langle I 0 L 0 \vert I^\prime 0 \right\rangle
\left\langle j {\frac{1}{2}} j^\prime {-\frac{1}{2}} \left| \right. L 
0  \right\rangle
\left\{
\begin{array}{ccc}
j' & j & L\\
I & I' & J\\
\end{array}
\right\} \ .
\end{eqnarray}

\bibliography{Algebraic_CC}

\end{document}